\documentclass[a4,]{scrartcl}
\usepackage{lmodern}
\usepackage{amssymb,amsmath}
\usepackage{ifxetex,ifluatex}
\usepackage{fixltx2e} 
\ifnum 0\ifxetex 1\fi\ifluatex 1\fi=0 
  \usepackage[T1]{fontenc}
  \usepackage[utf8]{inputenc}
\else 
  \ifxetex
    \usepackage{mathspec}
    \usepackage{xltxtra,xunicode}
  \else
    \usepackage{fontspec}
  \fi
  \defaultfontfeatures{Mapping=tex-text,Scale=MatchLowercase}
  
\fi
\IfFileExists{upquote.sty}{\usepackage{upquote}}{}
\IfFileExists{microtype.sty}{\usepackage{microtype}}{}
\usepackage[margin=2.5cm]{geometry}
\ifxetex
  \usepackage[setpagesize=false, 
              unicode=false, 
              xetex]{hyperref}
\else
  \usepackage[unicode=true]{hyperref}
\fi

\hypersetup{breaklinks=true,
            bookmarks=true,
            pdfauthor={},
            pdftitle={How much baseline correction do we need in ERP research? Extended GLM model can replace baseline correction while lifting its limits},
            colorlinks=true,
            citecolor=blue,
            urlcolor=blue,
            linkcolor=magenta,
            pdfborder={0 0 0}}
\usepackage{authblk}

\title{How much baseline correction do we need in ERP research? Extended GLM
model can replace baseline correction while lifting its limits}

\author[Max-Planck Institute for Psycholinguistics

P.O. Box 310

6500 AH Nijmegen,

The Netherlands]{
Phillip M. Alday 
}

\date{June 2019}
\usepackage{subcaption}
\usepackage{graphicx}

\begin{document}
\maketitle
\begin{abstract}
Baseline correction plays an important role in past and current
methodological debates in ERP research (e.g.~the Tanner v. Maess debate
in \emph{Journal of Neuroscience Methods}), serving as a potential
alternative to strong highpass filtering. However, the very assumptions
that underlie traditional baseline also undermine it, implying a
reduction in the signal-to-noise ratio. In other words, traditional
baseline correction is statistically unnecessary and even undesirable.
Including the baseline interval as a predictor in a GLM-based
statistical approach allows the data to determine how much baseline
correction is needed, including both full traditional and no baseline
correction as special cases. This reduces the amount of variance in the
residual error term and thus has the potential to increase statistical
power.
\end{abstract}

\hypertarget{introduction}{
\section{Introduction}\label{introduction}}

Baseline correction belongs to one of the standard procedures in ERP
research (cf. Luck 2005), yet comes with two inherent difficulties: the
choice of baseline interval and the assumption that there are no
systematic differences between conditions in the baseline interval.
Often discussed in conjunction with high-pass filtering, baseline
correction is argued to be an artifact free way to compensate for signal
drifts in electrophysiological recordings (cf.~the recent debate started
in the Journal of Neuroscience Methods: B. Maess, Schröger, and Widmann
2016; Tanner, Morgan-Short, and Luck 2015; Tanner et al. 2016; Burkhard
Maess, Schröger, and Widmann 2016; Widmann, Schröger, and Maess 2015).
In the following we will demonstrate that regardless of the choice of
baseline interval or highpass filter setting, traditional baseline
correction is never an optimal procedure with modern statistical
methods. In short the correct way to address potential bias introduced
by signal drifts is by including the baseline period in the statistical
analysis.

\hypertarget{the-general-linear-model-in-erp-research}{
\section{The General Linear Model in ERP
Research}\label{the-general-linear-model-in-erp-research}}

At the heart of all common analyses in ERP research, whether
repeated-measures ANOVA or various forms of explicit regression, is the
General Linear Model:

\[ \mathbf{y} = \sum_{i \in \text{covariates}} \beta_i \mathbf{x}_i + \epsilon \]
\[ \epsilon \sim{} N(0,\sigma^2)\]

where the \(\mathbf{y}\) represents a column vector of observed EEG data
(usually averaged over a given time-window and in ANOVA-based
approaches, averaged over trials), the \(\mathbf{x}_i\) are column
vectors of various predictors and covariates, the \(\beta_i\) represent
the (statistically determined) weights of the \(\mathbf{x}_i\), and
\(\epsilon\) represents the error term, i.e.~residuals, which are
assumed to be normally distributed. In its usual form, the error term is
assumed to be homogenous, i.e.~having the same variance across the
entire model and thus independent of any particular observation (the
\emph{homoskedacity} assumption). In the case of baseline-corrected
statistics analyses, we can decompose the \(\mathbf{y}\) column into

\[\mathbf{y} = \mathbf{y}_{\text{window}} - \mathbf{y}_{\text{baseline}}\]

(Note that it does not matter whether the baseline is subtracted from
the entire epoch before averaging within a given time window or
afterwards. This is because the baseline correction for a given epoch is
a constant and the average of the difference is the same as the
difference to the average.)

This means that we can re-express our GLM as:
\[ \mathbf{y}_{\text{window}} - \mathbf{y}_{\text{baseline}} = \sum_{i \in \text{covariates}} \beta_i \mathbf{x}_i + \epsilon \]

which we can further rewrite as:

\[ \mathbf{y}_{\text{window}} = \sum_{i \in \text{covariates}} \beta_i \mathbf{x}_i + \mathbf{y}_{\text{baseline}} + \epsilon \]

To highlight the fact that the baseline correction is now on the
``predictors'' side of the equation, we change its name from
\(\mathbf{y}_{\text{baseline}}\) to \(\mathbf{x}_{\text{baseline}}\)

\[ \mathbf{y}_{\text{window}} = \sum_{i \in \text{covariates}} \beta_i \mathbf{x}_i + \mathbf{x}_{\text{baseline}} + \epsilon \]

We note that this is just a special case of a linear model with the
baseline correction as a predictor, with the special case that
\(\beta_{\text{baseline}} = 1\) (and no baseline correction is exactly
the case that \(\beta_{\text{baseline}} = 0\)). This already suggests a
more general way forward: we make the baseline interval a proper
predictor and allow the model to determine the weight empirically.
Nonetheless, let us consider the usual assumptions of classical baseline
correction.

\hypertarget{the-underlying-assumptions-of-traditional-baseline-correction-make-it-irrelevant}{
\section{The Underlying Assumptions of Traditional Baseline Correction
Make It
Irrelevant}\label{the-underlying-assumptions-of-traditional-baseline-correction-make-it-irrelevant}}

For traditional baseline correction to be valid, we assume that
experimental conditions (whether traditional discrete, factorial
conditions or ``continuous conditions'' in more naturalistic and less
parametric designs) do not differ systematically in the
electrophysiological activity in their respective baseline intervals. If
they were to differ systematically in their baseline interval, then
traditional baseline correction would move effects from the baseline
window into window of interest (cf.~e.g. Luck 2005). Component overlap
between trials presents a particular set of problems for this assumption
(cf. Luck 2005), although component overlap within trials is also
problematic and several methods have been proposed to address this issue
(Smith and Kutas 2014b, 2014a). In the following, we will ignore this
particular problem for simplicity and without loss of generality.

As we have assumed no systematic differences in the baseline interval
between conditions, we can think of the vector of baseline values as
noise:
\(\mathbf{x}_{\text{baseline}} \sim{} N(\mu_{\text{baseline}},\sigma_{\text{baseline}}^2)\),
which we assume to be normally distributed without loss of generality.
In this case, our linear model simplifies to:

\[ \mathbf{y}_{\text{window}} = \sum_{i \in \text{covariates}} \beta_i \mathbf{x}_i +  \bar{\mathbf{x}}_\text{baseline} + \epsilon'\]

where \[ \epsilon' \sim{} N(0,\sigma^2 + \sigma_{\text{baseline}}^2) \]

In other words, under these assumptions, traditional baseline correction
increases the variance of the error term, i.e.~increases the noise,
without otherwise impacting the inferential engine beyond introducing a
shared offset \(\bar{\mathbf{x}}_\text{baseline}\), which will typically
be expressed as a change in the intercept term. However, we have made a
small yet potentially misleading equivalency, namely that ``no
systematic differences in the electrophysiological activity in the
baseline interval'' is the same as ``no systematic differences in the
baseline interval''. Other physical and environmental differences may
lead to conditions differing systematically in their baseline interval.
In the case that they differ only in their mean, then the previous
observation holds, although the offset introduced by the baseline is now
conditional on the experimental condition, i.e.~there is now an
interaction term with condition. If, however, the variance of the
baseline interval differs, then we no longer meet the assumption of
homoskedacity, as the resulting error term \(\epsilon'\) is not
homogenous across conditions.

We note at this point that the mathematics of traditional baseline
correction -- subtracting out a reference signal -- are the same as the
mathematics for re-referencing. It is no surprise then that baseline
correction suffers the same pitfalls as a bad reference, such as biasing
apparent topographies and introducing noise (cf. B. Maess, Schröger, and
Widmann 2016; Urbach and Kutas 2006). However, unlike re-referencing,
where each channel is shifted by the same time-dependent value and thus
the relative values remain the same even if the individual values change
(see Figures 1 and 4 in Lau et al. 2006 for an example), baseline
correction shifts each channel by a different time-independent signal
and can change the observed topography. As such, even more so than the
choice of reference, the choice of baseline influences the inferences
that can be made about observed effects (see Section
\ref{hpfilter_blwindow} below for further discussion on the choice of
baseline window).

Finally, this result also holds for analyses of spectral power (ERSP)
under the usual normalization procedure. In particular, the usual
normalization of dividing the target window by the baseline window and
then taking the logarithm of the quotient (i.e.~converting to decibels)
yields the same statistical model:
\[ \log \frac{ \mathbf{y}_{\text{window}} }{ \mathbf{x}_{\text{baseline}} } = \log \mathbf{y}_{\text{window}} - \log \mathbf{x}_{\text{baseline}}.\]
In the following, we will omit further explicit mention of
time-frequency analyses, but we note that all results and suggestions
apply equally to ERP and ERSP.

\hypertarget{explicit-regression-on-single-trial-data-as-an-optimal-solution}{
\section{Explicit Regression on Single-Trial Data as an Optimal
Solution}\label{explicit-regression-on-single-trial-data-as-an-optimal-solution}}

Returning to explicit regression without the baseline window included in
the error term, we can consider the simple case of one experimental
predictor:\footnote{Of course, in a real study, we would probably have
  multiple predictors, including topographic ones as well as random
  effects for e.g.~by-participant and by-item differences.}

\[ \mathbf{y}_{\text{window}} = \beta_0 + \beta_{\text{condition}}\mathbf{x}_{\text{condition}}  + \beta_{\text{baseline}}\mathbf{x}_{\text{baseline}} + \epsilon \]

In line with modern practice, we assume that this is a single-trial
analysis, although the same should hold, albeit less optimally, for
aggregated analyses. Including \(\mathbf{x}_{\text{baseline}}\) as a
predictor, we use the data to determine the weighting of the baseline
correction, with \(\beta_{\text{baseline}} = 1\) corresponding to
traditional baseline correction and \(\beta_{\text{baseline}} = 0\)
corresponding to no baseline correction. Now, if the conditions differ
in the amount of baseline correction ``necessary'', we can
straightforwardly address this by adding an interaction term to our
model:

\[ \mathbf{y}_{\text{window}} = \beta_0 + \beta_{\text{condition}}\mathbf{x}_{\text{condition}}  + \beta_{\text{baseline}}\mathbf{x}_{\text{baseline}} + \beta_{\text{condition,baseline}}\mathbf{x}_{\text{condition}}\mathbf{x}_{\text{baseline}} + \epsilon \]

This interaction term allows the amount of baseline correction to vary
by condition, as would be e.g.~necessary if changes in the external
environment during the experiment (electrode gel warming up, participant
sweating, changes in ambient electrical noise) change during the course
of experiment, especially for block designs. However, even in the case
of non-block designs, this actively accounts for issues resulting from
randomization order and can be complemented by added main-effect and
interaction terms for the trial sequence (or even smoother terms, cf.
Baayen et al. 2017; Tremblay and Newman 2015). As above for the main
effect, \(\beta_{\text{condition, baseline}} = 1\) corresponds to
traditional baseline correction where the average baseline correction
may vary by condition, while \(\beta_{\text{condition, baseline}} = 0\)
corresponds to a differential weighting of the necessity of baseline
correction by condition.

As this procedure allows the data to determine how much baseline
correction is warranted \emph{by condition}, it is optimal and not as
strongly dependent on the ``no systematic differences'' assumption. Like
GLM-based deconvolution methods, which model mixtures of time-lagged
influences on the signal, this technique reduces confounding by
explicitly modelling other influences on the signal, instead of mixing
them into the response. Moreover, this method includes traditional
baseline correction as well as no baseline correction as special cases
and thus supersedes those methods. As noted above, this result holds
equally well for single-trial time-frequency data under the usual
normalization procedure.

The notion of confound is also useful for a more intuitive derivation of
the optimality of this approach, where baseline is a covariate. Baseline
correction is not there to create a true zero per se, but rather as an
inferential control (cf. Urbach and Kutas 2006). As we noted previously,
good experimental design can and should also function as a way for
inferential control, and indeed the usual baseline assumptions
correspond exactly to a particular method of experimental control.
However, a more general and more powerful technique is to adjust for
potential confounds statistically, by including potential confounds as a
covariate. Rather than making a priori assumptions about the impact of
the confound, this procedure allows for determining its actual influence
and allows for a broader class of experimental designs where the
confound cannot be controlled via systematic manipulation or
experimental procedure (Sassenhagen and Alday 2016).

This method can also be viewed as a computationally simple special case
of regression methods such as rERP (Smith and Kutas 2014b, 2014a),
without lagged predictors and marginalized over distinct time windows.
The method presented here has the advantage that it fits much more
easily into existing computational and statistical frameworks, trivially
works with modern mixed-effects models for simultaneously modelling both
participant and item variance (Baayen, Davidson, and Bates 2008; Clark
1973; Judd, Westfall, and Kenny 2012), and is no more expensive
computationally than other contemporary methods (see worked example
below). Finally, this method also subsumes and generalizes other
baseline-normalization methods such as traditional baseline correction,
especially when interactions with the baseline predictor are included.

This method does, however, have a few ``disadvantages''. It functions
best with unaggregated (i.e.~single-trial) data and explicit regression
approaches (i.e.~not AN(C)OVA); however, these are considered best
practice anyway (for the general statistical preference for explicit
estimation, see (Cumming 2014; Kruschke and Liddell 2017), for insights
gleaned from single-trial analyses of ERP data, see e.g. (Frömer, Maier,
and Rahman 2018; Gaspar, Rousselet, and Pernet 2011; Pernet, Sajda, and
Rousselet 2011; Hauk et al. 2006), for the advantages of a multi-level
regression approach using mixed-effects models, see (Baayen, Davidson,
and Bates 2008; Clark 1973; Judd, Westfall, and Kenny 2012)).
Numerically, other issues may arise if there is large signal drift and
thus variables on vastly different scales; however, once again best
statistical practice, namely centering and scaling variables, provides a
solution to this problem.\footnote{This is sometimes addressed as part
  of the signal processing, via a special case of baseline correction,
  namely subtracting the mean of the whole trial from each trial.
  However, as the activity between conditions is assumed to differ
  between trials, this again violates the assumptions of baseline
  correction and can introduce effects into other time windows. This is
  especially problematic for large-amplitude and/or prolonged effects.
  While not problematic for statistical analyses carefully focused on a
  single time window of interest, this is still less optimal than simply
  scaling the variables in the regression model.} More challenging is
that the additional parameters in these models increase both
computational complexity and the amount of data necessary for reliable
parameter estimation. This is especially true for models including
topographical information (e.g.~channel name or position in a
multi-channel recording). The computational complexity is hard to
address, but the requirement for more data is again in line with
contemporary best practice to address the chronic lack of power in
neuroscience (cf. Button et al. 2013; Szucs and Ioannidis 2017).
Regularization (e.g.~ridge regression or LASSO in the frequentist
framework, sparsity priors in the Bayesian framework) can also help.
This method is also somewhat more difficult to integrate into procedures
not based on the GLM, such as independent-component analysis and source
localization, although probably not prohibitively so. For example, this
technique would provide an interesting way to improve stationarity and
thus potentially enhance IC decompositions of epoched data without
depending on the strong filters often used in such contexts.\footnote{We
  are indebted to a helpful reviewer for suggesting this approach.}
Finally, this method does not completely address issues related to the
selection of the baseline interval, which remains an open question and a
researcher degree of freedom, but some general guidelines are suggested
in the next section.

\hypertarget{hpfilter_blwindow}{
\section{Relationship to Highpass Filtering and Choice of Baseline
Window}\label{hpfilter_blwindow}}

It is common to refer to baseline correction as an alternative or
complementary to (strong) highpass filtering. However, baseline
correction can also be interpreted as a highpass filter in its own right
(albeit an unusual one). In intuitive terms, baseline correction removes
the changes in the signal between epochs and can thus be interpreted as
removing slow drifts and thus low-frequency components. Like a filter,
baseline correction, both traditional and regression-based, also has a
free parameters that influence its effect on the data.

All things equal, a longer baseline window will tend to be less noisy or
variable compared to a shorter one. In statistical terms, a longer
baseline window corresponds to a larger sample drawn from a random
variable and will thus tend to offer a better estimate of its true mean
with less variance (i.e.~both more accurate and more precise). However,
all things are rarely equal and longer baseline windows present
additional difficulties: they require longer interstimulus intervals
(potentially disruptively long ones for many research questions) and/or
potentially include parts of the evoked response from the previous
stimulus, thus changing the meaning of reference point for later evoked
potentials. This suggests that a baseline window on the order of a few
hundred milliseconds may be the sweet spot for many experimental designs
under typical laboratory conditions without large high-frequency
artifacts (see empirical example below for a brief comparison of
different baseline windows).

Beyond the length of the baseline window, the relative position of the
baseline window to the time-locking events and critical events is also
important. Because the position of the baseline window within an epoch
is \emph{absolute} and not relative compared to a given sample (as in a
typical filter), baseline correction will generally not remove slow
drifts within an epoch. In traditional baseline correction, the entire
epoch is shifted by a constant offset and thus the overal slope is not
affected: translations are shape-preserving transforms. In the
regression-based correction proposed here, the drift away from the
calibration given by baseline will eventually lead to the baseline
weight shrinking to zero. This is unsurprising in the sense that a
distant baseline window is a poor baseline window (e.g.~the first 2
seconds of EEG recording are not used as the baseline window for all
trials in that recording). As such, baseline correction is not a
substitute for but rather a complement to traditional highpass
filtering.

The choice of baseline window should also be shaped by the research
question. The logic of baseline correction, as highlighted by Urbach and
Kutas (2006), is not to establish a true zero (which may or may not be
meaningful for a measure such as voltage that is inherently a
difference) but rather a meaningful reference point or control with
which to compare successive changes and thereby infer causality. For a
classical pre-stimulus baseline, the event-related potential thus shows
the change in the electric field following the stimulus (or, more
generally, event of interest): the state after the stimulus relative to
the (average) state before. For a baseline consisting of the entire
epoch, the event-related potentials show the change in the electric
field relative to its average of a time interval which includes the
event of interest. This does not show as directly that the state
afterwards is different than the state before and instead only shows the
difference to the average state. When the difference to the average
state is larger after an event of interest than the difference to the
average state before the event of interest, then this can also be taken
as indirect evidence of event-related change; however, this second stage
of ``difference of differences'' is implicitly an additional baseline
correction to the prestimulus interval. The no-baseline-correction case
corresponds to assuming that the reference point aligns with true zero,
which may be a reasonable assumption e.g.~for studies with stronger
highpass filtering and comparable stimulation before the critical event.
The advantage of using regression-weighted baseline correction is that
the data determines the evidence that the chosen reference point
(baseline window) differs from true zero and how to weight its
contribution because the reference point itself is a noisy measurement.
In other words, using a deterministic baseline is ignoring the error
bars on the control given by the baseline window.

This is crucial when interpreting topographies. Traditional baseline
correction necessarily projects the inverse scalp topography into the
epoch, but the weighting in the regression-based approach properly
controls for scalp topography instead of forcibly shifting it. This is
achieved in two ways. First, the weighting of the baseline window can
differ by electrodes. Second, the weighting of the baseline window can
differ by condition. In either case, this can be achieved by performing
the baseline correction on each electrode or condition separately (as in
traditional baseline correction) or by including topographical position
or condition as interaction effects in the regression model (for a
pooled estimate). By applying such proper statistical control, we can
avoid many of the biases that lie at the heart of Urbach and Kutas's
arguments. The empirical example in the next section shows how
traditional baseline correction can be misleading in such cases, but the
regression-based approach properly controls for topographical
differences in the baseline conditions.

In brief, baseline correction serves a similar role to highpass
filtering and suffers many of the same potential pitfalls in terms of
artifacts, both causal and acausal. Moreover, each has a number of
similar tradeoffs: longer baseline intervals and stronger highpass
filters better correct for some types of noise in the signal but at
increased risk of additional artifacts. However, one does not completely
replace the other and the combined choice of baseline window and
highpass filter should reflect the tradeoffs necessary for a particular
experimental design. Regression-based baseline correction supersedes
traditional baseline correction, but does not eliminate the need for
appropriate highpass filtering.

\hypertarget{empirical-example-n400-paradigm-with-environmental-noise}{
\section{Empirical Example: N400 Paradigm with Environmental
Noise}\label{empirical-example-n400-paradigm-with-environmental-noise}}

In the following, we aim to demonstrate the claims above via an
empirical example. We re-analyze data from Tromp et al. (2017), a
classical semantic mismatch N400 paradigm, but conducted in virtual
reality with a cross-modal mismatch. The virtual reality setting
presents a particular challenge because of the potential for
environmental noise and movement artifacts. Such noise and artifacts
could potentially cause signal changes despite no violation of the ``no
systematic differences in electrophyiological activity'' assumption and
thus necessitate a correction for signal drift.

Using MNE-Python v0.17.1 (Gramfort et al. 2013) and in line with the
original analysis, (continuous, non-epoched) data were re-referenced to
the linked mastoids and bandpass filtered from 0.1 to 40 Hz (passband
edge; zero-phase FIR filter with a Hamming window and
\texttt{fir\_design=\textquotesingle{}firwin\textquotesingle{}}, all
other parameters left as
\texttt{\textquotesingle{}auto\textquotesingle{}}). These filter
settings should eliminate line-noise and very slow drifts without
inducing problematic artifacts, but traditional wisdom suggests that
they do not eliminate the need for baseline correction (cf. Tanner et
al. 2016). As in the original analysis, the baseline interval consisted
of the 100ms immediately before (auditory) stimulus onset. Analyses
conducted with other high-pass filter edges (0.1, 0.3, 0.5, 1.0 Hz) are
presented below in summary form for comparison, but are not discussed at
depth nor further analyzed. Trials with instantaneous amplitude
exceeding ±75 µV were excluded from further analysis. Although Tromp and
colleagues analyzed both the N400 time window and an earlier time
window, we restrict ourselves to their N400 window (350--600 ms).

It is important to note that the original data were recorded at 500 Hz
and filtered online with a lowpass filter at 200 Hz and a highpass
filter at 0.016 Hz. The file metadata show that the highpass filter was
applied both in hardware and in software, while the lowpass filter was
applied only in software. Although Tromp and colleagues originally
reported online highpass filtering at 0.01 Hz, Brain Products amplifiers
specify their cutoff in time (here: 10s) and the corresponding frequency
cutoff is calculated as \(1/2\pi{}t\) following analog filter convention
and not as \(1/t\) as is common in other areas. As such, the raw data
already reflect two forward passes of a weak highpass filter. This will
of course greatly attenuate the sorts of drift that baseline correction
serves to correct, but is not an unusual recording setup and as such
demonstrates that the role baseline correction plays under actual
laboratory conditions.

All analysis source code as well as the pre-processed single-trial data
are available on OpenScience Framework (https://osf.io/pnaku/). There
are data for each of the above filter settings as well as for several
different baseline windows (500ms pre-stimulus, 200ms pre-stimulus,
100ms pre-stimulus, 200ms post-stimulus, average across entire epoch).
It is beyond the scope of this manuscript to discuss all possible
combinations of baseline interval and filter settings in depth, but we
do briefly examine the impact of the baseline interval for the primary
highpass filter setting (0.1 Hz) below.

\hypertarget{differences-in-the-baseline-are-illusory}{
\subsection{Differences in the Baseline are
Illusory}\label{differences-in-the-baseline-are-illusory}}

\begin{figure}[!htb]
  \centering
  \begin{subfigure}[]{0.3\textwidth}
    \centering
    \includegraphics[width=\textwidth]{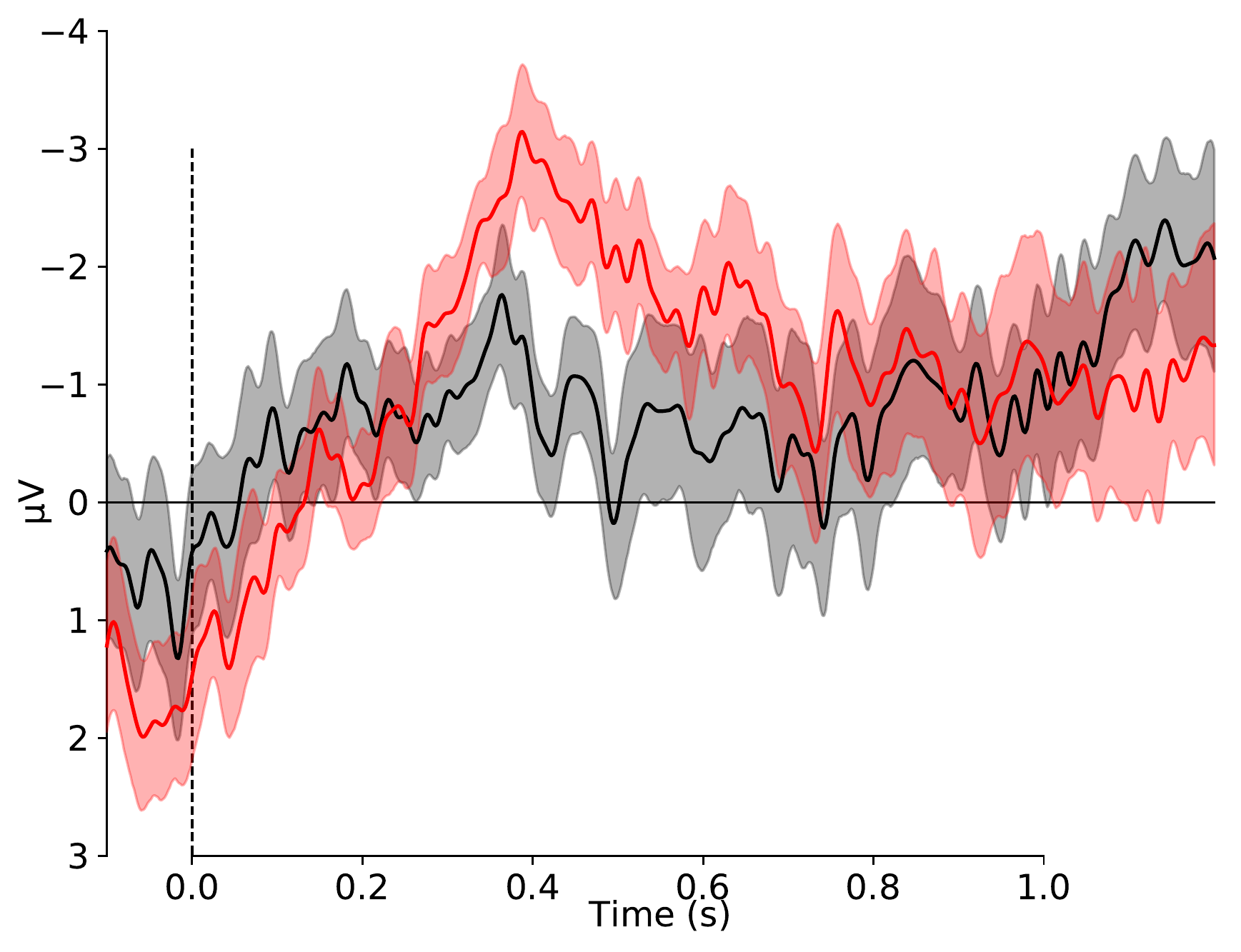}
    \caption{No baseline correction}
    \label{erp:nobaseline}
  \end{subfigure}
  \quad
  \begin{subfigure}[]{0.3\textwidth}
    \centering
    \includegraphics[width=\textwidth]{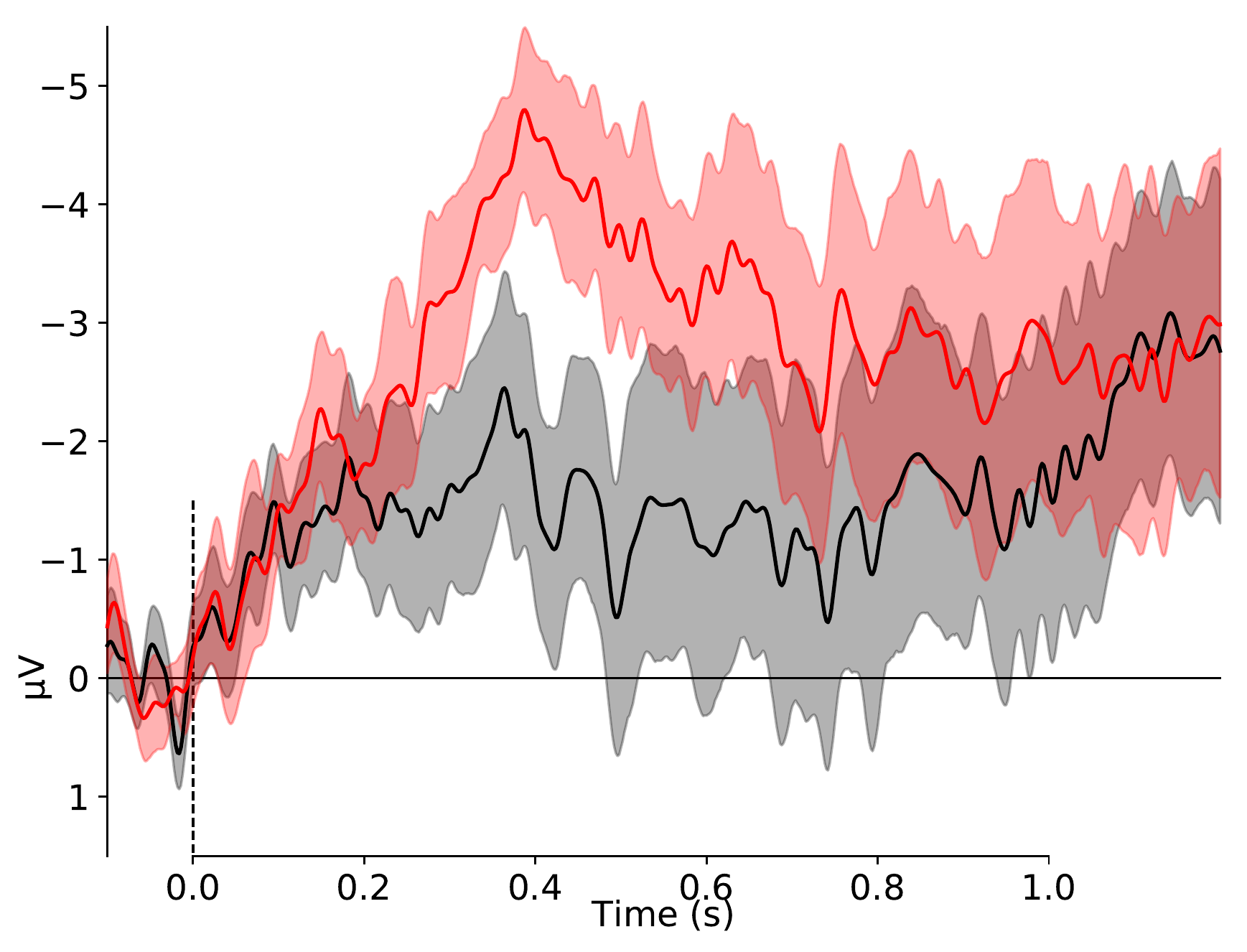}
    \caption{Traditional baseline}
    \label{erp:baseline}
  \end{subfigure}
  \quad
    \begin{subfigure}[]{0.3\textwidth}
    \centering
    \includegraphics[width=\textwidth]{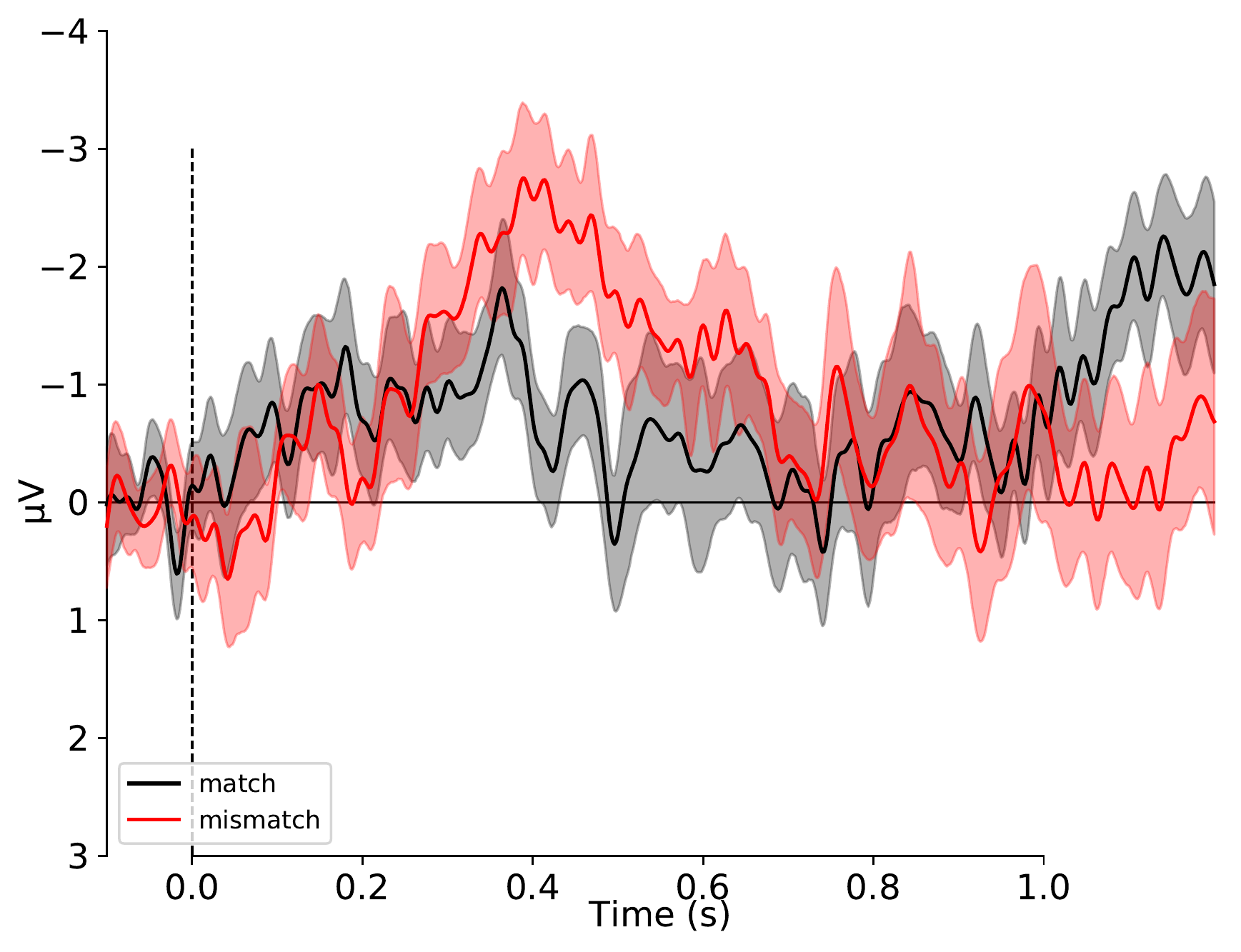}
    \caption{Regression-based baseline}
    \label{erp:regression}
  \end{subfigure}
  \caption{Comparison of baseline-correction strategies for waveforms at the apex electrode (Cz).
           Left, no baseline correction.
           Center, traditional baseline correction.
           Right, regression-based baseline correction.
           Solid lines represent grand averages computed over subjects.
           Shaded areas indicate 83\% bootstrap confidence intervals; non-overlap at this level is equivalent to zero lying outside of the 95\% confidence interval of the difference.
           Although the vertical midpoint of the subplots differs, the overall vertical span is the same across all subplots, so that visual size of the confidence intervals is directly comparable.
           The noise introduced by deterministically combining the baseline window and its associated variance with the rest of the timecourse leads to the broader confidence intervals in the center plot.
           See Figure \ref{erpdiff:all}, for a presentation of the difference waves with a shared vertical midpoint.
           For the regression-based baseline correction, the plotted time course was computed using a linear-model at each timepoint and electrode with main and interaction effects of baseline and condition. This is equivalent to computing the GLM as above but with a 1-sample time window and repeating this for all samples.}
  \label{erp:all}
\end{figure}

\begin{figure}[!htb]
  \centering
  \begin{subfigure}[]{0.3\textwidth}
    \centering
    \includegraphics[width=\textwidth]{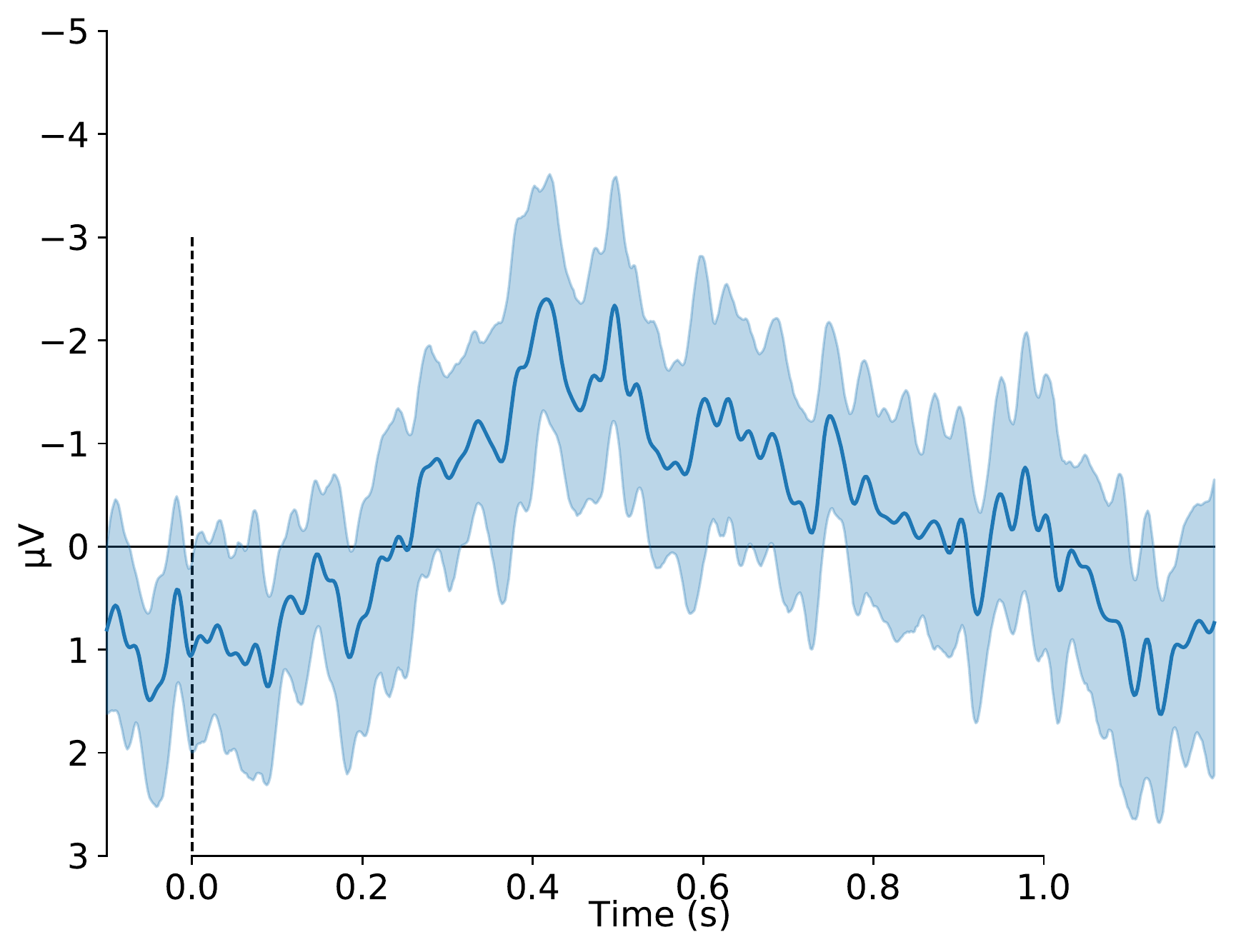}
    \caption{No baseline correction}
    \label{erpdiff:nobaseline}
  \end{subfigure}
  \quad
  \begin{subfigure}[]{0.3\textwidth}
    \centering
    \includegraphics[width=\textwidth]{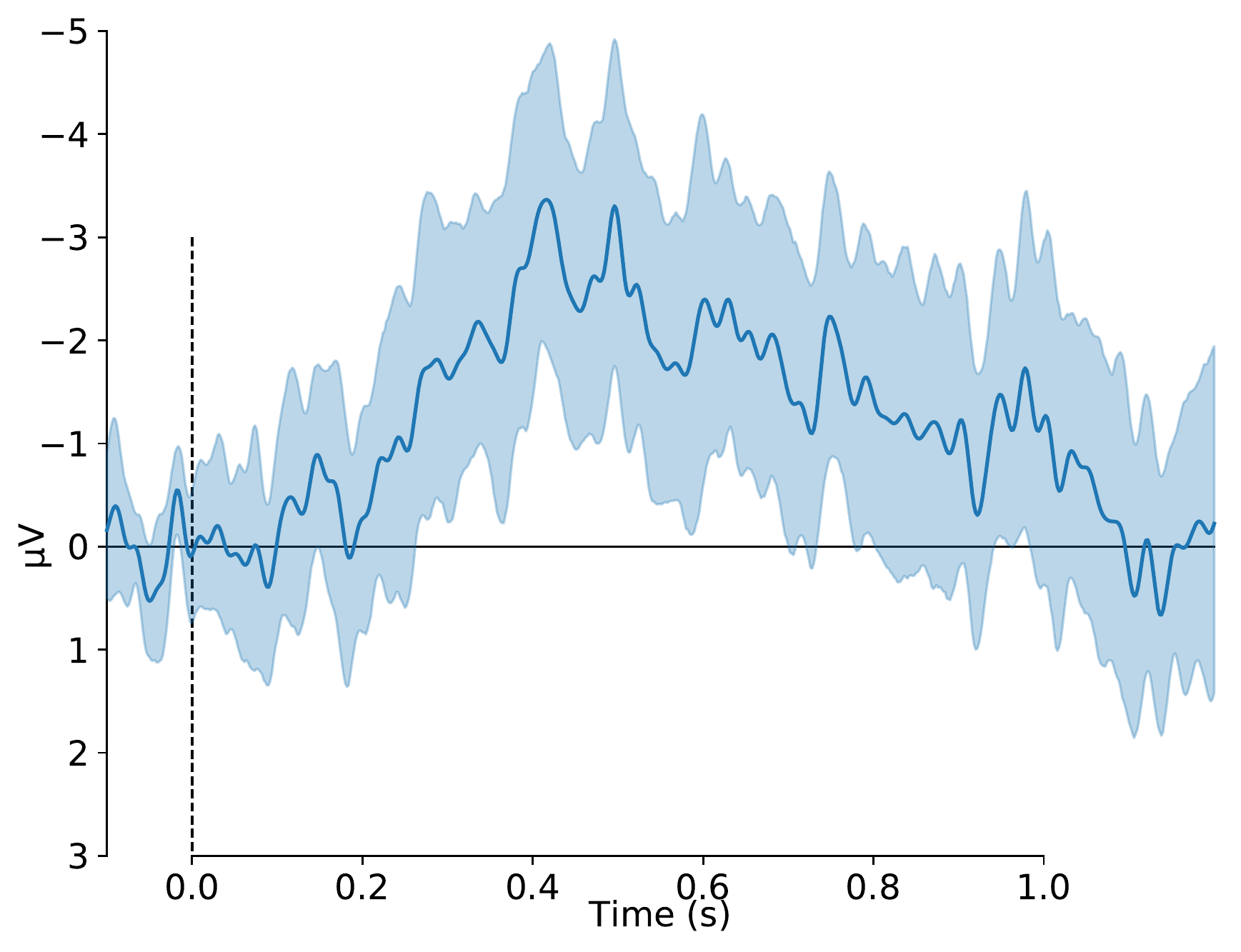}
    \caption{Traditional baseline}
    \label{erpdiff:baseline}
  \end{subfigure}
  \quad
    \begin{subfigure}[]{0.3\textwidth}
    \centering
    \includegraphics[width=\textwidth]{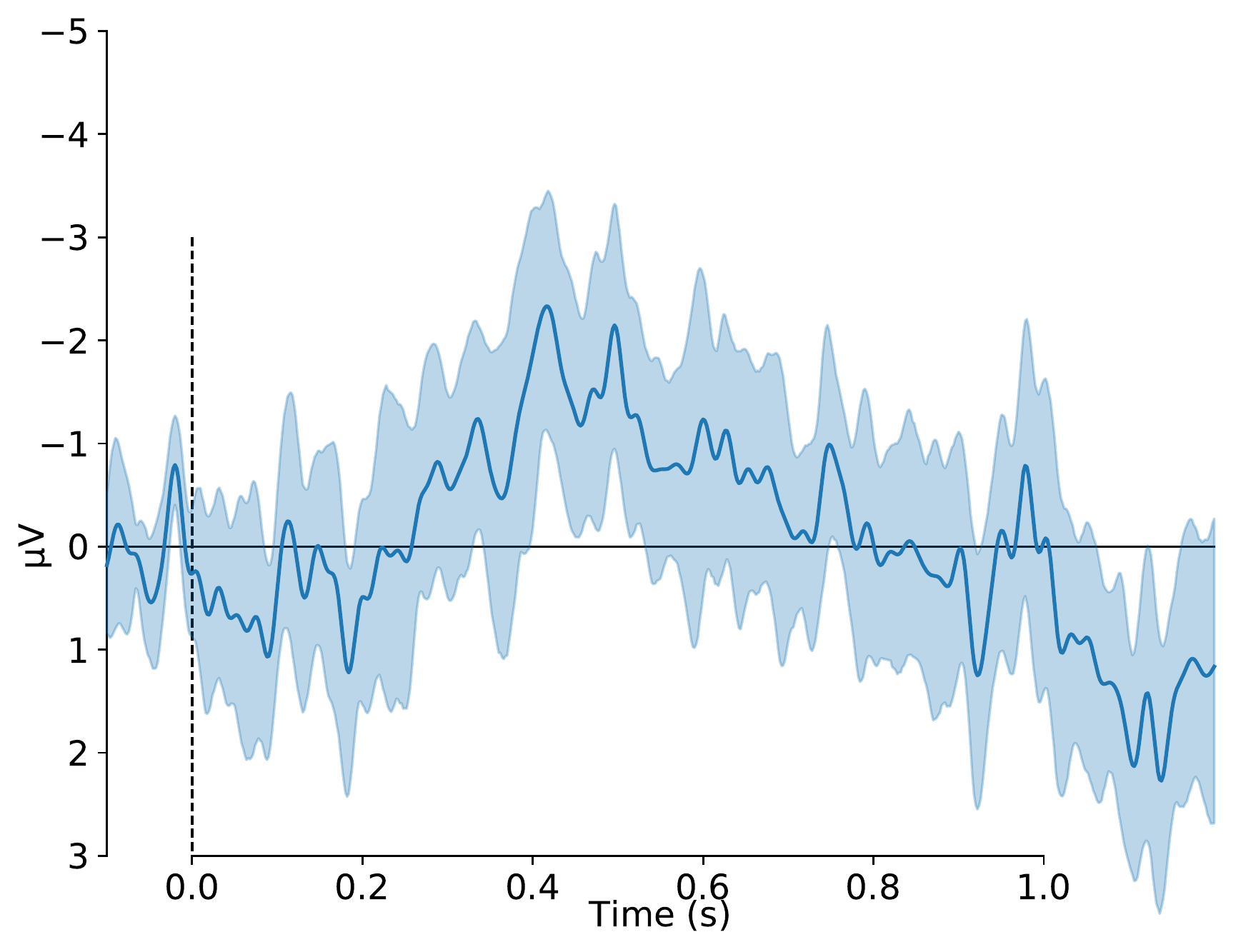}
    \caption{Regression-based baseline}
    \label{erpdiff:regression}
  \end{subfigure}
  \caption{Comparison of baseline-correction strategies for difference waves at the apex electrode (Cz).
           Left, no baseline correction.
           Center, traditional baseline correction.
           Right, regression-based baseline correction.
           Solid lines represent grand averages computed over subjects.
           Shaded areas indicate 95\% bootstrap confidence intervals.
           In contrast to Figure \ref{erp:all}, all subplots share a single vertical midpoint.
           The baseline correction was computed for each condition and applied before the computation of the difference waves.
           For the regression-based baseline correction, the plotted time course was computed using a linear-model at each timepoint and electrode with main and interaction effects of baseline and condition.
           The difference waves were computed at the subject level and then averaged.}
  \label{erpdiff:all}
\end{figure}

The grand-average waveforms from the apex electrode Cz are presented in
Figure \ref{erp:all} with the corresponding difference waves in Figure
\ref{erpdiff:all}. Figure \ref{erp:nobaseline} presents the waveform
without any baseline correction. Although the grand averages themselves
look distinct in the baseline window, we see that the confidence
intervals overlap and correspondingly the confidence interval for the
difference wave crosses zero (Figure \ref{erpdiff:nobaseline}). We are
thus unable to reject the null hypothesis that the observed difference
between conditions in the baseline interval occurred by chance alone.
Less rigorously, the waveforms are statistically indistinguishable in
the baseline window and the apparent differences in the baseline window
are not distinguishable from noise. Figure \ref{erp:baseline} presents
the waveform with traditional baseline correction. We note how the
confidence intervals become broader; moreover, there is an apparent, yet
misleading prolonged separation of the waveforms well beyond the N400
time window, which is also apparent in the difference wave in Figure
\ref{erpdiff:baseline}. Again, the overlap in the confidence intervals
suggest this difference is not distinguishable from noise; however, this
distinction would be lost in typical ERP plots without confidence
intervals. Finally, Figure \ref{erp:regression} presents the
regression-based baseline strategy applied to each timepoint. We see
that the confidence intervals are much narrower than in the traditional
baseline correction. Moreover, the overall time course of the N400
effect is much more apparent and much more temporally focal in the
difference plot (Figure \ref{erpdiff:regression}).

\begin{figure}[!htb]
  \centering
  \begin{subfigure}[]{0.5\textwidth}
    \centering
    \includegraphics[width=\textwidth]{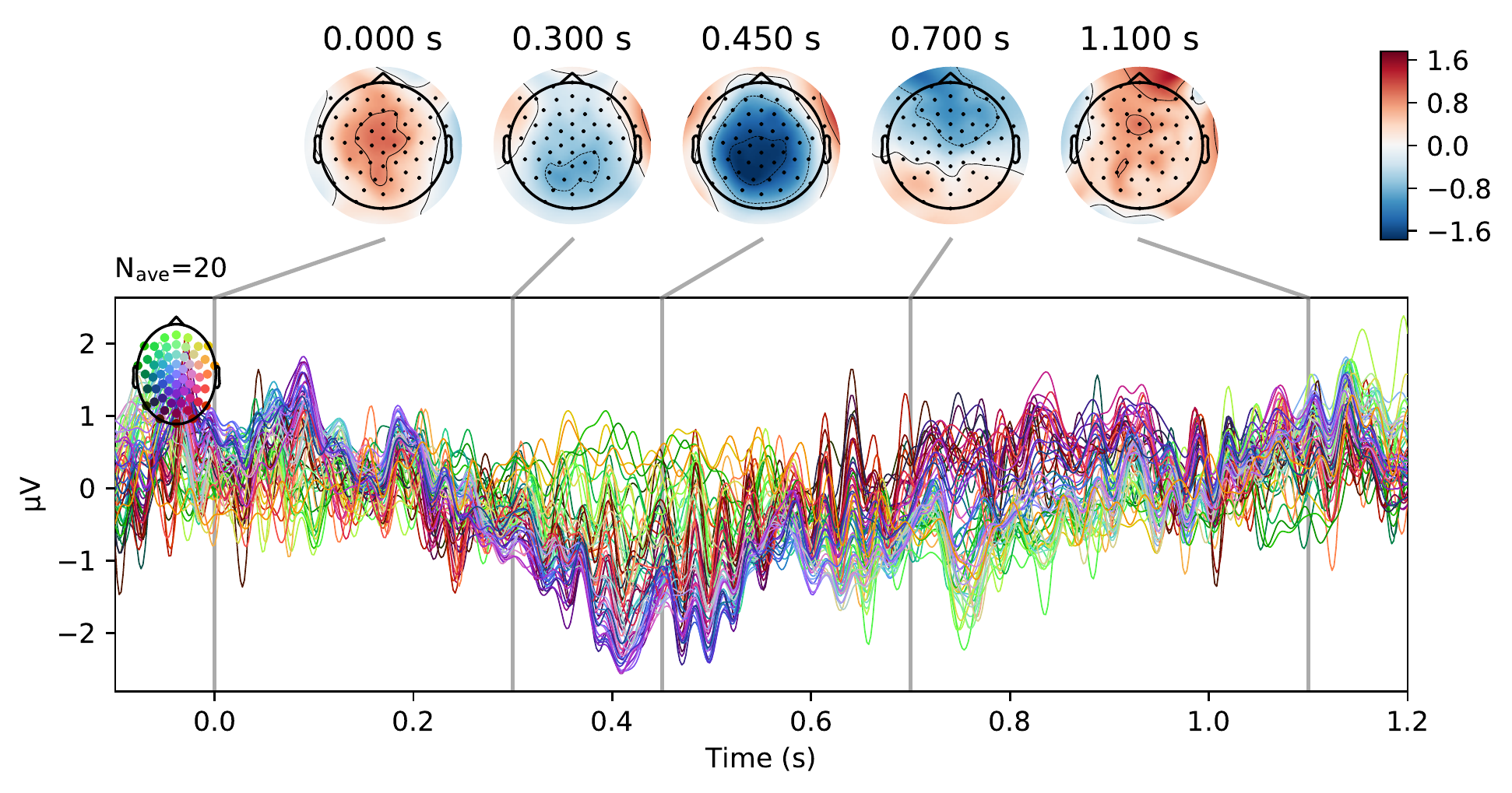}
    \caption{No baseline correction}
    \label{joint:nobaseline}
  \end{subfigure}

  \begin{subfigure}[]{0.5\textwidth}
    \centering
    \includegraphics[width=\textwidth]{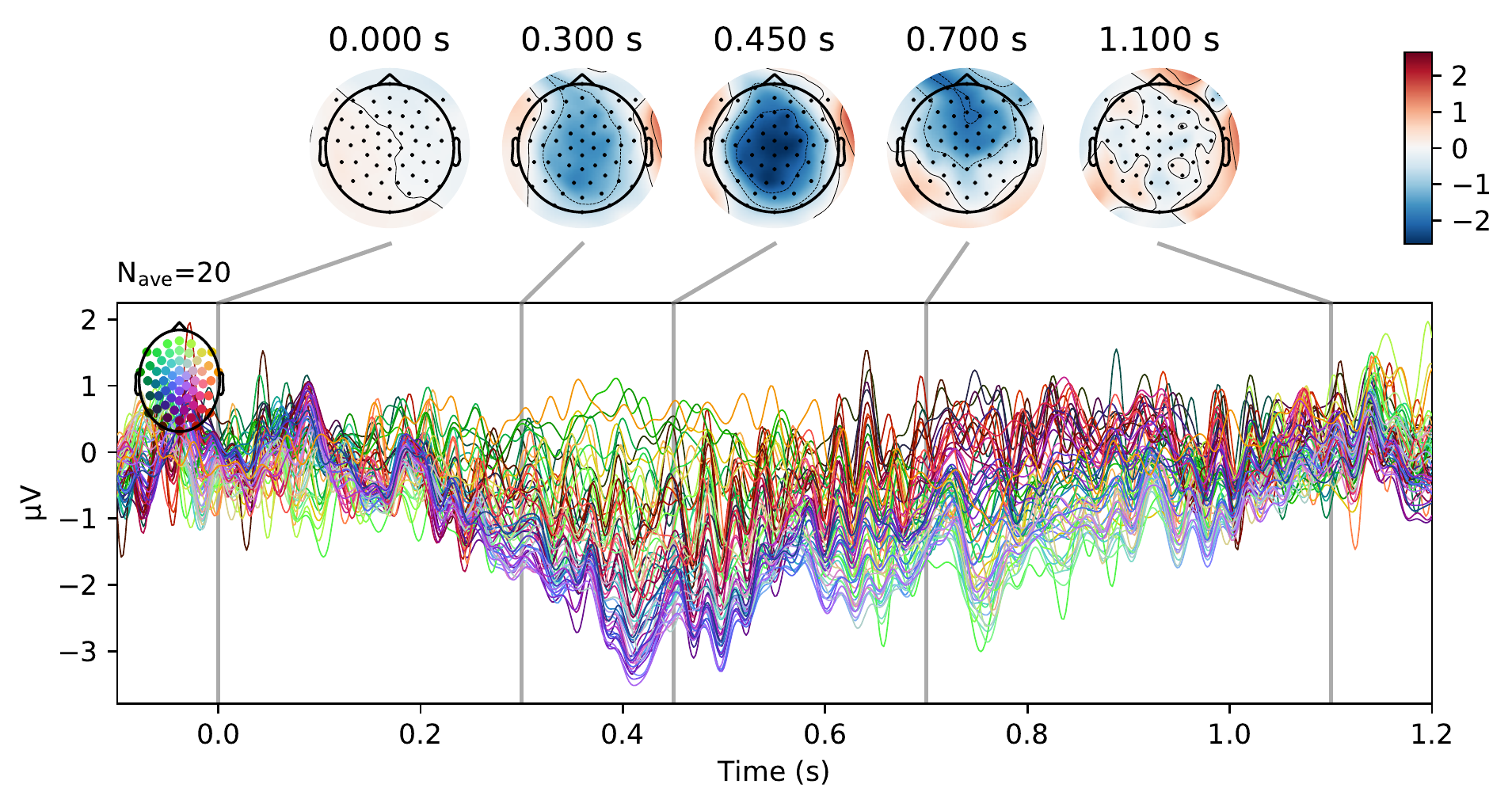}
    \caption{Traditional baseline}
    \label{joint:baseline}
  \end{subfigure}

   \begin{subfigure}[]{0.5\textwidth}
    \centering
    \includegraphics[width=\textwidth]{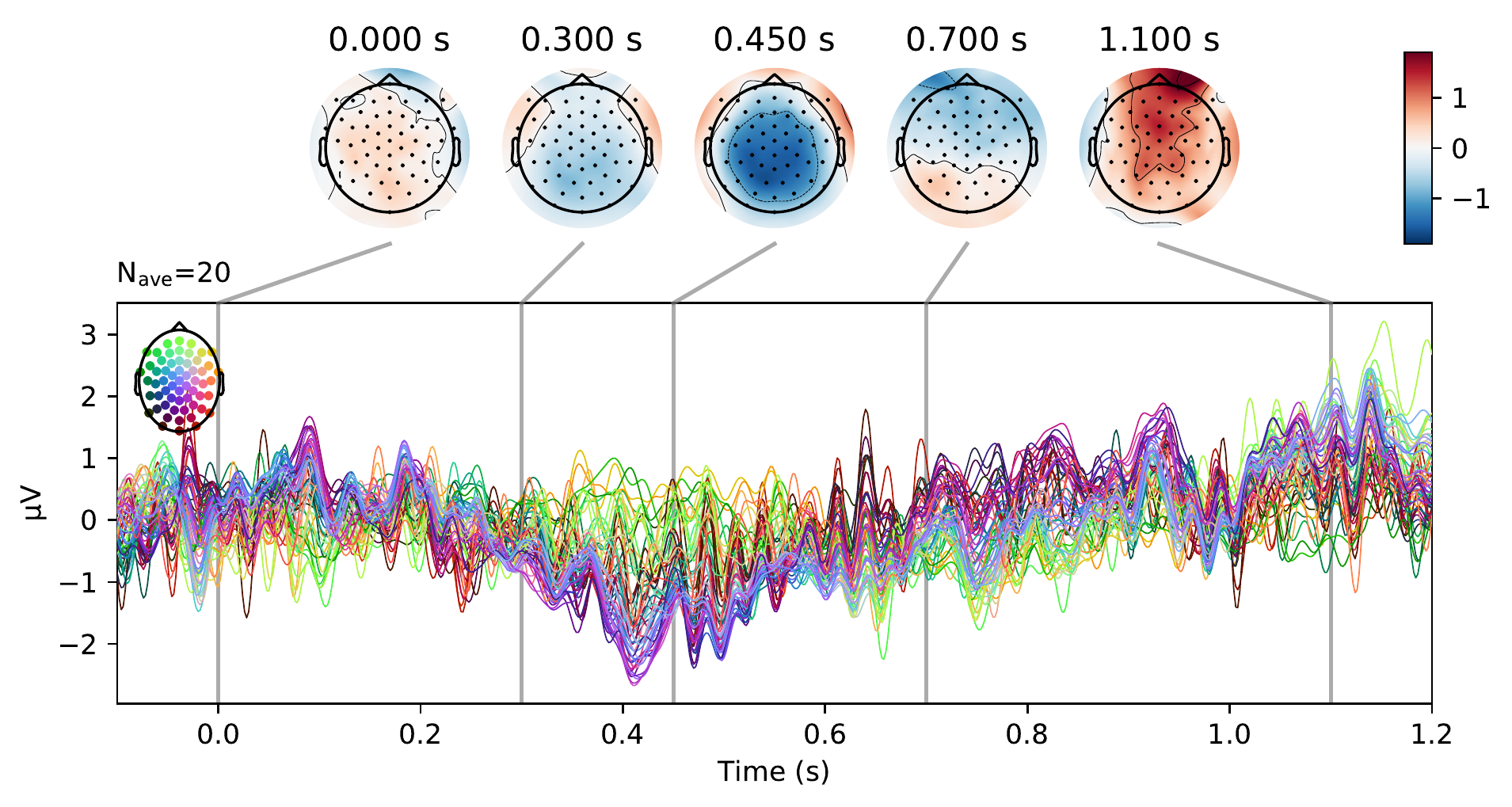}
    \caption{Regression-based baseline}
    \label{joint:regression}
  \end{subfigure}
  \caption{Comparison of baseline-correction strategies for the topography of difference waves.
           Top, no baseline correction.
           Center, traditional baseline correction.
           Bottom, regression-based baseline correction.
           Each colored line represents a single channel.
           The topographic plots depict average activity in the marked time ±100ms.
           For the regression-based baseline correction, the plotted time course was computed using a linear-model at each timepoint and electrode with main and interaction effects of baseline and condition.
           The baseline correction was computed for each condition and applied before the computation of the difference waves.
           The difference waves were computed at the subject level and then averaged.}
  \label{joint:all}
\end{figure}

Figure \ref{joint:all} displays the topography of the grand-average
difference waves. Note that the overall topography does not change
greatly between baseline-correction methods for this experiment because
stimulation before onset of the critical item was comparable. Taken
together, Figures \ref{erpdiff:all} and \ref{joint:all} suggest that
regression-based baseline correction reduces the size of the N400
effect. This is not quite accurate; instead, traditional baseline
correction leads to a slight overestimation of the size of the N400
effect. This is discussed in more depth below.

\begin{figure}[!htb]
\centering
\includegraphics[width=\textwidth]{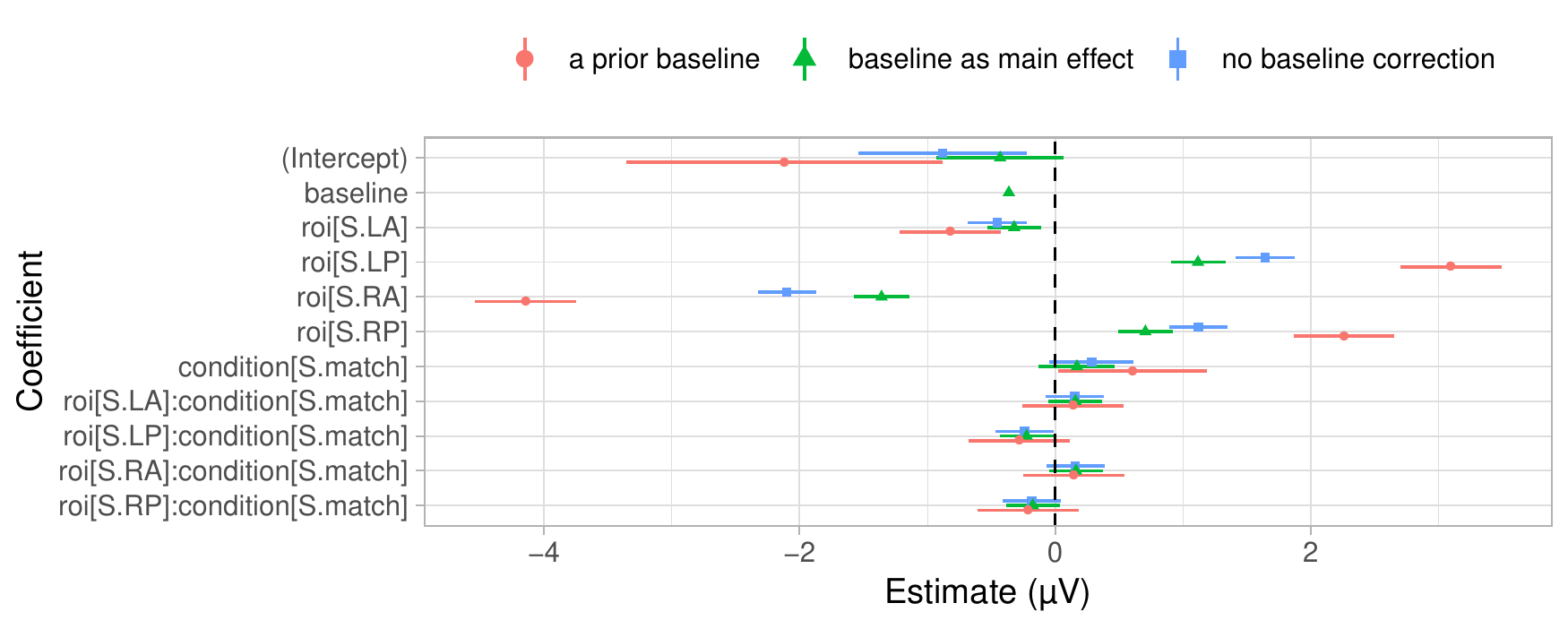}
\caption{Coefficient plot comparing estimates from different baseline correction strategies in the later time window (600--800ms). Intervals are 95\% profile confidence intervals. Note the extremely small, yet extremely precise estimate for the (effect of the) baseline window. Model selection preferred the regression model including baseline as a main effect but not further interacting with ROI or condition; see below for a more extensive example using the primary window of interest (350--500ms). The impact of the small bias introduced by traditional baseline correction is apparent in the confidence interval for the effect of condition -- even a small change downwards would have lead to a rejection of the null hypothesis.}
\label{fig:coef:late}
\end{figure}

Most interestingly, the later N400 effect around 600-800ms reported by
Tromp and colleagues with traditional baseline correction has a
different topography than the early one around 300-500ms.\footnote{I am
  indebted to a helpful reviewer for pointing out this shift in
  topography and positing that it may be a baseline artifact.} While
they reported no overall interaction between condition and topography
within each time window, they did not compare topographies between time
windows. Meanwhile, both no baseline correction and regression-based
baseline correction suggest an extremely weak effect near zero across
the entire scalp (Figures
\ref{joint:nobaseline},\ref{joint:regression}). Examining the topography
in the baseline window given in the no-baseline-correction plot (Figure
\ref{joint:nobaseline}), we see that traditional baseline correction
projects this small albeit non-significant difference in means forward
in time, where it combines with the minimal effect present in that time
window to generate the larger observed effect with its distorted
topography and duration. This graphical impression is supported by
analysis with mixed-effects models: the inclusion of the baseline in the
model improves fit and removes the effect of condition (Figure
\ref{fig:coef:late}). In contrast, the primary N400 effect in the
350--600ms time window retains its topography across correction methods.

For all of these plots, we note that the the bootstrap confidence
intervals computed sample-wise per electrode on single-subject averages
do not correspond directly to the statistics used in the analysis below.
In particular, the analyses below include subject and item variance
simultaneously and are computed on trialwise window and ROI averages.
The window and ROI averaging will generally increase the signal-to-noise
ratio, and as has been noted many times (e.g. Clark 1973; Baayen,
Davidson, and Bates 2008; Judd, Westfall, and Kenny 2012), item variance
cannot be ignored, especially in language studies. This is apparent
below (Table \ref{tbl:baseline.mod}), where the between-item variance is
larger than the between-subject variance.

\hypertarget{the-pre-stimulus-baselines-influence-on-later-components-is-not-what-you-think}{
\subsection{The Pre-Stimulus Baseline's Influence on Later Components is
Not What You
Think}\label{the-pre-stimulus-baselines-influence-on-later-components-is-not-what-you-think}}

\begin{figure}[!htb]
\centering
\includegraphics[width=0.5\textwidth]{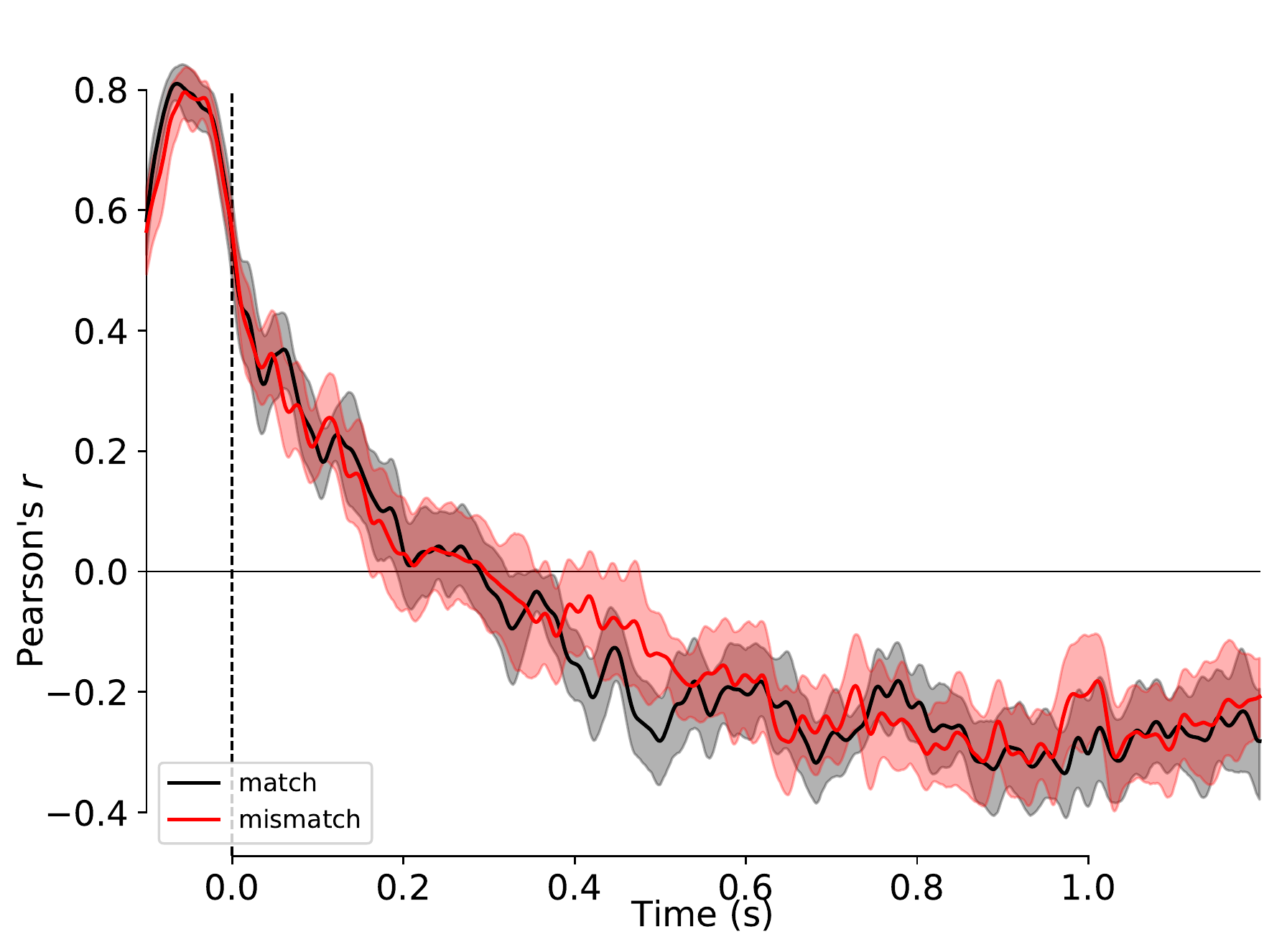}
\caption{Correlation of electrophysiological signal with baseline interval at the apex electrode (Cz).
         Pearson's $r$ was computed over single-subject averages; shaded areas indicate bootstrapped 95\% confidence intervals. }
\label{fig:corr}
\end{figure}

The misleading duration and amplitude of the N400 effect in the plot
with traditional baseline correction is partly the result of traditional
baseline correction's ability to bias later components \emph{in the
wrong direction}. Figure \ref{fig:corr} shows the correlation of the ERP
for each condition with the baseline interval over time. Unsurprisingly,
the correlation with the mean of the baseline interval is quite high
within the baseline interval; however, the correlation drops off
rapidly, reaching zero less than 200ms after stimulus onset. Somewhat
disconcertingly, the correlation in typical N400 and P600 time windows
is non-zero and \emph{negative}. This suggests that traditional baseline
correction is shifting the waveform in the wrong direction. As voltage
is inherently a relative measure, this bias, shared amongst all
conditions, is not particularly problematic per se. Nonetheless, the low
magnitude of the correlation at larger latencies indicates that there is
little shared covariance between the baseline window and the target
window. In other words, applying traditional \emph{a priori} baseline
correction fails to correct bias introduced by the baseline. Moreover,
traditional baseline correction may introduce additional bias and will
necessarily introduce the additional variance from the baseline
interval. This suggests that the baseline interval is most relevant for
the early exogenous, perceptual components. Again, including the
baseline interval as a predictor in the statistical model applies the
correct amount of correction as determined by the data -- and that level
of correction is expected to differ between datasets. For example, DC
recordings without online or offline highpass filtering will necessarily
require more correction than those such as here with both online and
offline highpass filters.

\begin{figure}[!htb]
\centering
\includegraphics[width=\textwidth]{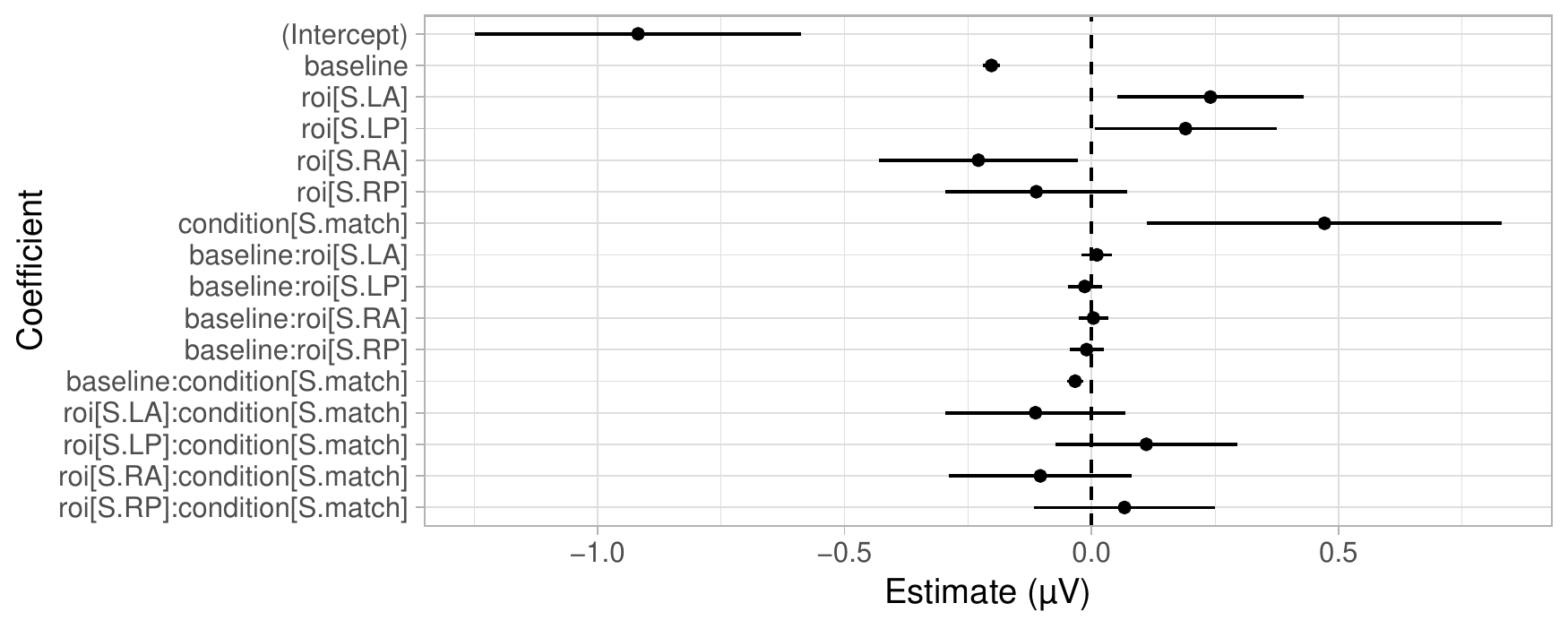}
\caption{Coefficient plot for the model presented in Table \ref{tbl:baseline.mod}. Intervals are 95\% profile confidence intervals. Note the extremely small, yet extremely precise estimate for the (effect of the) baseline window.}
\label{fig:coef}
\end{figure}

We see this in the mixed-effects model for the N400 time window,
presented in Table \ref{tbl:baseline.mod} and Figure \ref{fig:coef}.
Although the main effect for the baseline window has a large
\(t\)-value, the actual size of the effect is quite small and in the
wrong direction. (Recall from above that traditional baseline correction
corresponds to a regression weight of \(+1\)). We also note that there
is an interaction of baseline with condition, which traditional baseline
correction could not have accommodated.

\begin{table}[!htb]
\caption{Summary of full model with pairwise interactions between topography, manipulation and the baseline. All categorical contrasts are sum-coded. ROIs are named by laterality (L vs R) and sagitality (A vs P) or the midline (M). Model fitted with \texttt{lme4} version 1.1.20 (Bates et al., 2015).}
\label{tbl:baseline.mod}
Linear mixed model fit by maximum likelihood 

\bigskip
\begin{tabular}{ccccc}
 AIC & BIC & logLik & deviance & df.resid \\ 
 40623 & 40782 & -20289 & 40577 & 7187 \\ 
  \end{tabular}

\bigskip
Scaled residuals:

\begin{tabular}{ccccc}
 Min & 1Q & Median & 3Q & Max \\ 
 -5.44 & -0.64 & 0.01 & 0.64 & 4.42 \\ 
  \end{tabular}

\bigskip
Random effects:

\begin{tabular}{llll}
 Groups & Term & Std.Dev. & Corr \\ 
 item & (Intercept) & 1.02305 &  \\ 
   & condition[S.match] & 0.68514 & -0.366 \\ 
  subj & (Intercept) & 0.48629 &  \\ 
   & condition[S.match] & 0.67955 & 0.228 \\ 
  Residual &  & 3.94258 &  \\ 
  \end{tabular}

Number of obs: 7210, groups:  item, 80; subj, 20.

\bigskip
Fixed effects:

\begin{tabular}{rrrr}
  & Estimate & Std. Error & t value \\ 
 (Intercept) & $-$0.92 & 0.17 & $-$5.6 \\ 
  baseline & $-$0.2 & 0.0088 & $-$23 \\ 
  roi[S.LA] & 0.24 & 0.096 & 2.5 \\ 
  roi[S.LP] & 0.19 & 0.094 &   2 \\ 
  roi[S.RA] & $-$0.23 & 0.1 & $-$2.2 \\ 
  roi[S.RP] & $-$0.11 & 0.094 & $-$1.2 \\ 
  condition[S.match] & 0.47 & 0.18 & 2.7 \\ 
  baseline:roi[S.LA] & 0.011 & 0.016 & 0.7 \\ 
  baseline:roi[S.LP] & $-$0.014 & 0.018 & $-$0.77 \\ 
  baseline:roi[S.RA] & 0.0039 & 0.015 & 0.25 \\ 
  baseline:roi[S.RP] & $-$0.01 & 0.018 & $-$0.57 \\ 
  baseline:condition[S.match] & $-$0.033 & 0.0087 & $-$3.8 \\ 
  roi[S.LA]:condition[S.match] & $-$0.11 & 0.093 & $-$1.2 \\ 
  roi[S.LP]:condition[S.match] & 0.11 & 0.094 & 1.2 \\ 
  roi[S.RA]:condition[S.match] & $-$0.1 & 0.095 & $-$1.1 \\ 
  roi[S.RP]:condition[S.match] & 0.067 & 0.094 & 0.71 \\ 
  \end{tabular}

\end{table}
 
\hypertarget{model-complexity-and-fit-and-their-impact-on-statistical-power}{
\subsection{Model Complexity and Fit and Their Impact on Statistical
Power}\label{model-complexity-and-fit-and-their-impact-on-statistical-power}}

While the model presented in Table \ref{tbl:baseline.mod} may seem much
more complex to fit and interpret than a model without the baseline
predictors, this is not the case. As elsewhere in statistics, we can
include additional covariates as controls without further interpreting
those covariates. In other words: we can safely ignore the terms related
to baseline correction, but we cannot omit them from the model. As
reflected in the shifted vertical midpoint in Figure \ref{erp:all}, the
baseline term will will have an impact if we compute e.g.~marginal
means, but that does not preclude us from interpreting the effects
attributable purely to our experimental manipulation. Moreover, if the
interpretation of the interaction between the baseline and the
experimental manipulation is of interest, then it is no different than
the interpretation of the interaction between topographical predictors
and the experimental manipulation.

\begin{figure}[!htb]
\centering
\includegraphics[width=\textwidth]{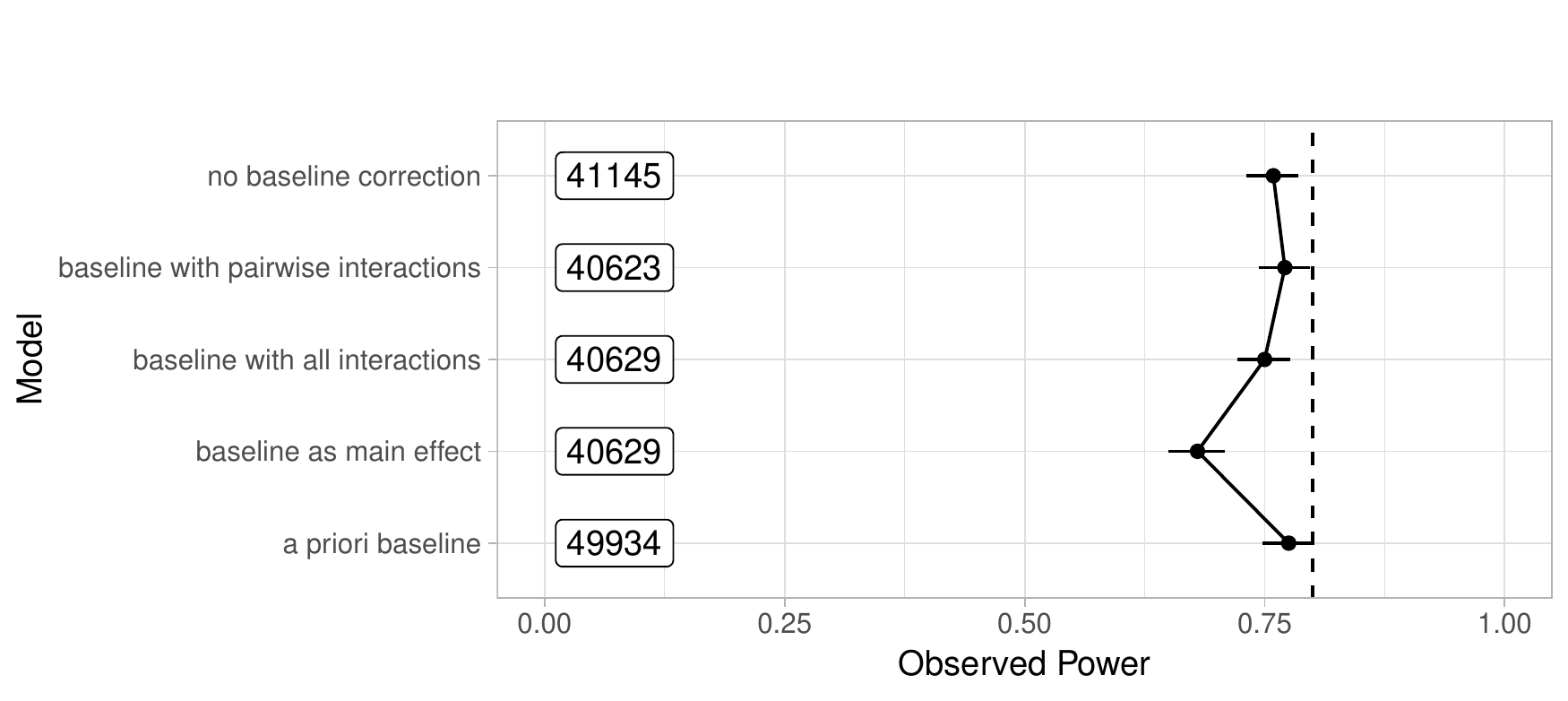}
\caption{Statistical power and model fit for different types of baseline correction.
         Boxed numbers on the left-hand are AIC (Akaike Information Criterion; Akaike, 1974), a combined measure of model fit and model parsimony.
         Smaller AIC is better; a 5-point difference is often considered significant.
         Intervals represent 95\% confidence intervals.
         The dashed vertical line indicates the traditional 80\% power threshold.
         Power computed from  1000 simulations with the \texttt{simr} package (Green and Macleod, 2016).}
\label{fig:power}
\end{figure}

Following the ongoing debate about the tradeoffs in Type-I error, power
and model complexity (e.g. Matuschek et al. 2017; Barr et al. 2013), we
can consider the impact of additional predictors on model fit and
statistical power. Figure \ref{fig:power} shows that the improved fit
resulting from including the baseline window as a predictor more than
compensates for the potential loss in power from the additional
predictors. Moreover, the reduced variance in the dependent variable
results in faster convergence of the numerical optimization procedure
and thus computation time is also not worsened by the additional model
complexity. For this particular dataset, the models with additional
terms for the interaction of the baseline with condition and topography
do show an improved fit (as measured by log likelihood), but the
accompanying increase in model complexity exceeds the corresponding
improvement in model fit when comparing the pairwise interaction model
to the full interaction model (\(\Delta\text{AIC}\)= 6, and
corresponding likelihood-ratio test \(\chi^2(4)=2.8\), \(p=0.6\)). We
therefore prefer the more parsimonous pairwise interaction model over
the full interaction model. Crucially, the model with traditional,
\emph{a priori} baseline correction performs the worst in terms of model
fit. The minimal apparent increase power is thus irrelevant because a
poorly fitting model calls the overall validity of inference into
question. We see here empirically what we demonstrated mathematically
above: traditional baseline correction reduces power and biases our
inferences.

\begin{figure}[!htb]
\centering
\includegraphics[width=\textwidth]{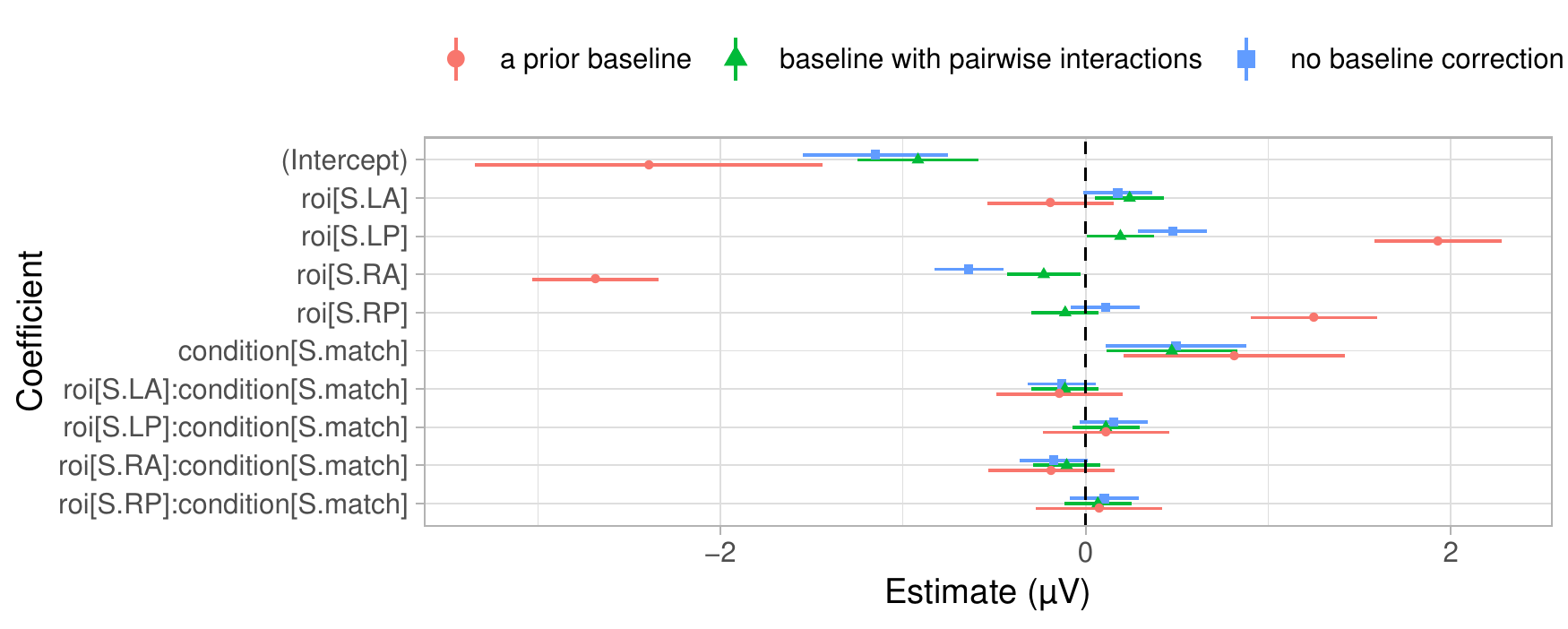}
\caption{Coefficient plot comparing the estimates from the models corresponding to different baseline strategies. Intervals are 95\% profile confidence intervals. Note the much larger confidence intervals for the \emph{a priori} baseline, but otherwise overall similar pattern of effects for the experimental manipulation and its topography. The differences in main effects in topography are an example of the topographical biases inherent in traditional baseline correction and reflect the combined topography of the baseline interval and average topography across both conditions, while the interaction model separates these effects.}
\label{fig:modelcmp}
\end{figure}

For comparison, the estimates from \emph{a priori} baseline, no
baseline, and the pairwise model are plotted in Figure
\ref{fig:modelcmp} (see also Figure \ref{fig:coef:late} for a similar
comparison in the late N400 time window examined in a post-hoc analysis
by Tromp and colleagues). Overall the pattern of effects is similar
across models, except that the model with the \emph{a priori} baseline
has much larger estimates and larger confidence intervals. For the main
effects of topography, this reflects the topographical biases inherent
in traditional baseline correction and reflect the combined topography
of the baseline interval and average topography across both conditions,
while the interaction model separates these effects.

The larger estimate for the experimental manipulation also leads to its
high power estimate (cf.~Figure \ref{fig:power}). Although its
confidence interval is much broader than the other models, the mean
value is higher and so the lower-edge of the confidence is further away
from zero. This in turns leads to higher observed power, which is known
to be biased in this way (cf. Hoenig and Heisey 2001; Gelman and Carlin
2014).

\hypertarget{choice-of-baseline-window-and-highpass-filter}{
\subsection{Choice of Baseline Window and Highpass
Filter}\label{choice-of-baseline-window-and-highpass-filter}}

\begin{figure}[!htb]
\centering
\includegraphics[width=\textwidth]{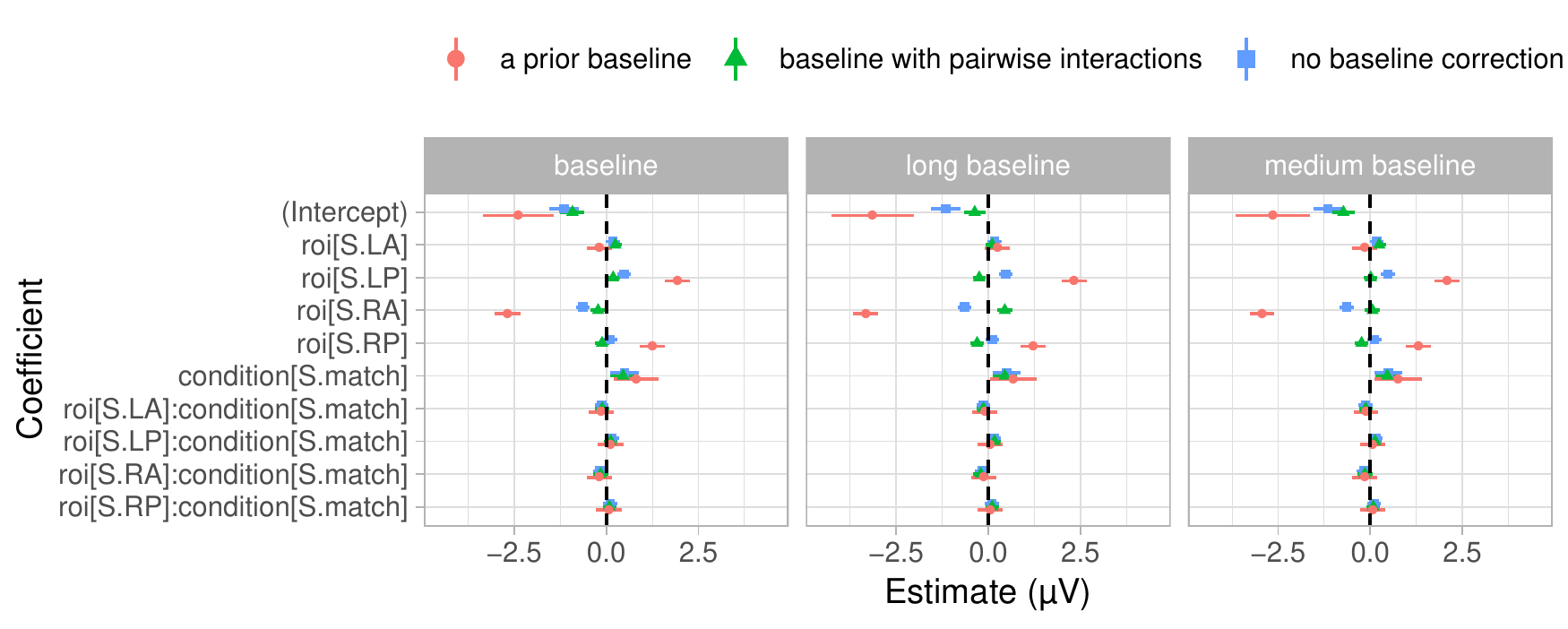}
\caption{Coefficient plot comparing the estimates from the models corresponding to different baseline strategies with different baseline windows. Intervals are 95\% profile confidence intervals. The long baseline corresponds to 500ms prestimulus, the medium to 200ms prestimulus and the `default' baseline to 100ms prestimulus. The different baseline strategies and windows were estimated with separate models.}
\label{fig:blwin:effects}
\end{figure}

\begin{figure}[!htb]
\centering
\includegraphics[width=\textwidth]{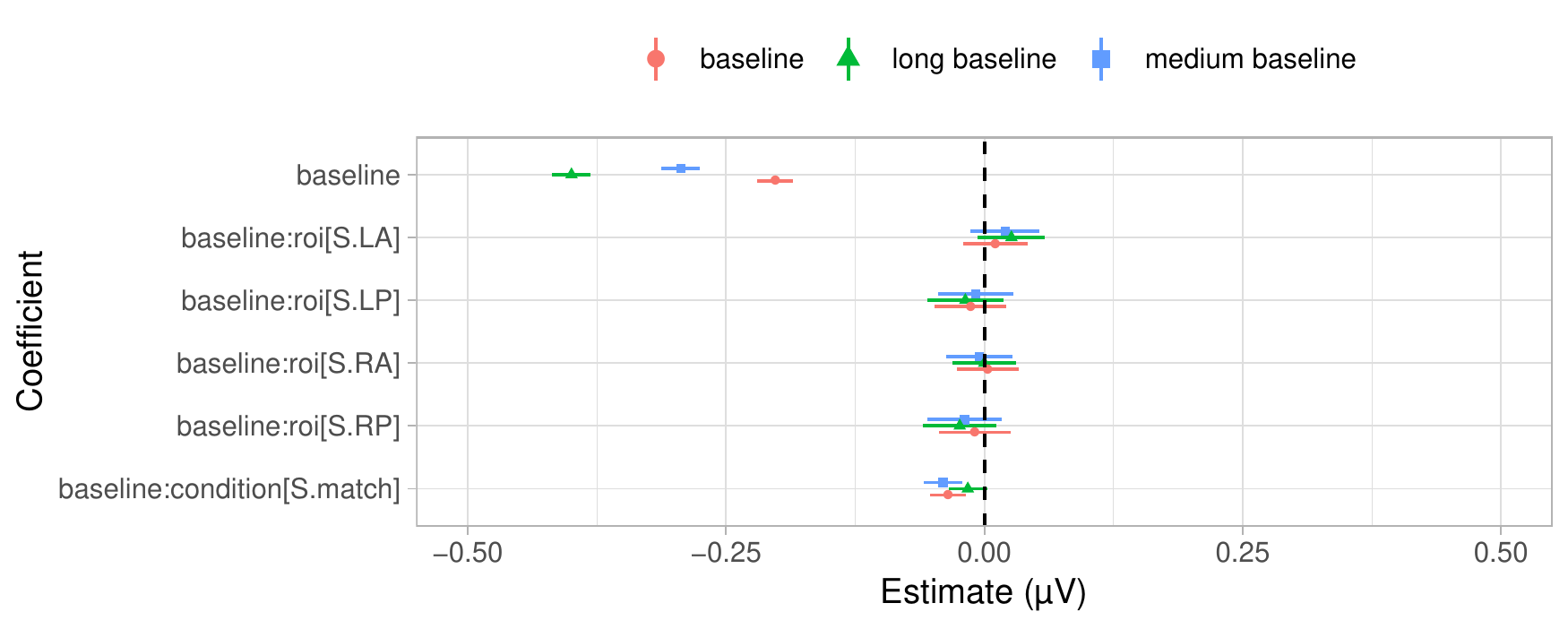}
\caption{Coefficient plot comparing the estimated weights awarded to different baseline intervals. Intervals are 95\% profile confidence intervals. The long baseline corresponds to 500ms prestimulus, the medium to 200ms prestimulus and the `default' baseline to 100ms prestimulus. The different baseline windows were estimated with separate models, all including pairwise interactions of the baseline interval and other predictors.}
\label{fig:blwin:weight}
\end{figure}

The baseline window and highpass filter used in the analysis thus far
were chosen to match the original analysis by Tromp and colleagues.
Given the overall experimental design and considerations on the impact
of baseline window discussed above (Section \ref{hpfilter_blwindow}), we
do not expect a large difference for longer pre-stimulus windows. We
tested this empirically by computing the same pairwise interaction model
for the same baseline (100ms prestimulus), a long baseline (500ms
prestimulus) and a medium baseline (200ms prestimulus). As Figure
\ref{fig:blwin:effects} shows, the overall pattern of effects, both
between and within models, did not change between conditions although
the absolute magnitude of the intercept term (reflecting the average
voltage across all conditions and ROIs) did change. Similarly, the
weight awarded to the baseline window changed (see Figure
\ref{fig:blwin:weight}), but its interactions with ROIs and condition
did not (reflecting an overall matching of the baseline pre-stimulus
interval across conditions). The change in both the intercept and weight
of the baseline term reflect a change in the absolute voltage measured
in the N400 window, but the absolute change on a relative scale is less
interesting than the impact it has on the estimate of the effect of
interest, which was minimal: the estimates for condition and its
topographical interactions did not differ much between baseline windows.
Note that different experimental designs with different stimulation and
noise constraints can lead to a longer or shorter baseline being
preferable, but for this experiment, the choice of prestimulus baseline
window did not have a huge impact.

\begin{figure}[!htb]
\centering
\includegraphics[width=\textwidth]{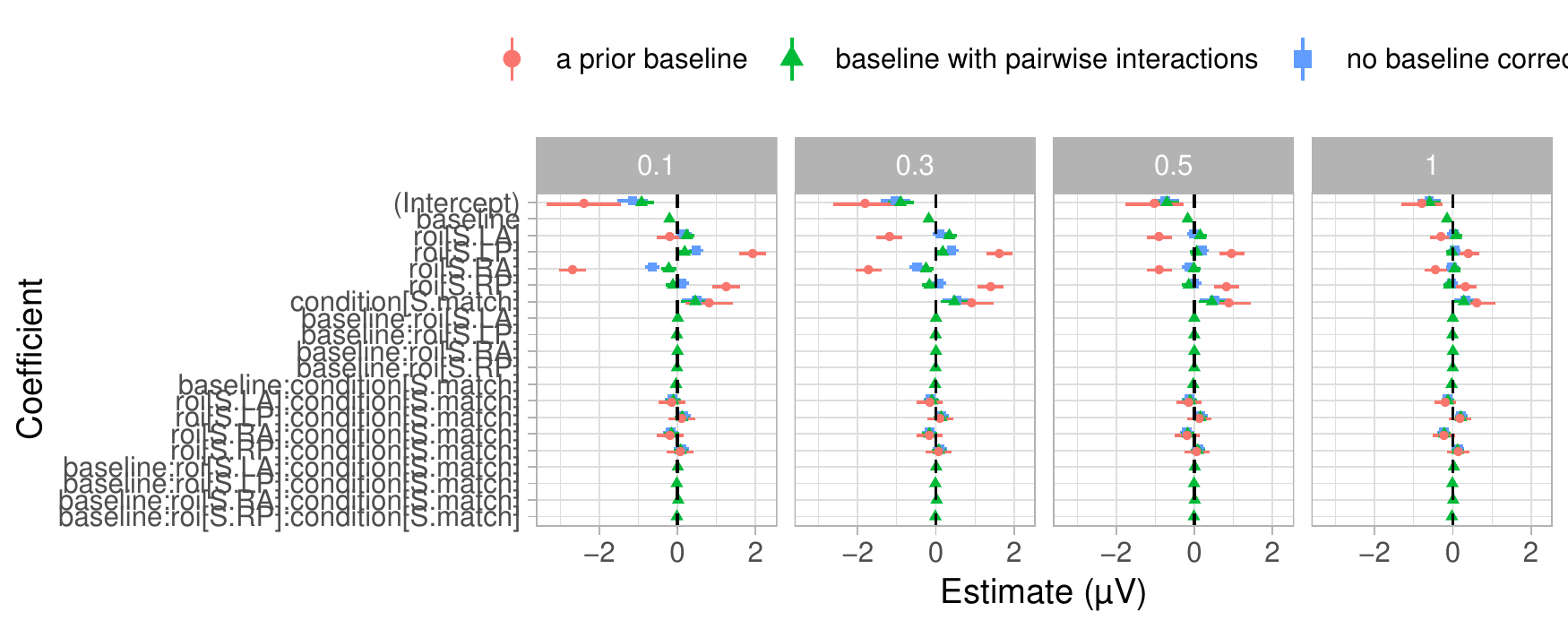}
\caption{Coefficient plot comparing the estimates from the models corresponding to different baseline strategies and different highpass filter settings. Intervals are 95\% profile confidence intervals.  All filters are bandpass zero-phase FIR filters with an upper passband edge of 40 Hz and a lower passband edge corresponding to the value in the plot. The different baseline strategies and filter settings were estimated with separate models. Note stronger filtering shrinks all effects to zero but those attributable to drift (intercept, baseline, topographical main effects) and not the targeted experimental manipulation (condition and its interactions) more strongly.}
\label{fig:hp}
\end{figure}

Similarly, the choice of high-pass filter did not greatly impact the
effect of interest here, as seen in Figure \ref{fig:hp}. All filter
settings with the exception of the relevant passband edge were the same
as above (zero-phase FIR) and for simplicity these models were only
computed using the 100ms prestimulus window as the baseline correction.
Stronger filtering shrinks all effects towards zero but those
attributable to drift (intercept, baseline, topographical main effects)
more strongly than the fast changes due to targeted experimental
manipulation (condition and its interactions). In addition to the
potential to shrink events of interest to zero with strong enough
filtering, filters can also introduce other artifacts not obvious in the
statistical models, as discussed at length in the Tanner-Maess debate.
Moreover, passband edge is not the only relevant filter setting -- the
choice of causal vs.~acausal, zero-phase or not, filter-length and IIR
vs.~FIR -- all involve a number of tradeoffs whose scope exceeds the
present manuscript.

\hypertarget{bayesian-analysis}{
\subsection{Bayesian Analysis}\label{bayesian-analysis}}

Despite the theoretical and empirical evidence presented above, some
researchers may still have a strong \emph{a priori} belief in the
necessity of the traditional baseline procedure. To that end, we again
note that the data-driven, model-based approach presented here will
yield traditional baseline correction, \emph{when the data support it}.
Moreover, we can accommodate our \emph{a priori} beliefs as part of
statistical model. Using the R package \texttt{brms} (version
2.7.0 , Bürkner 2017) to inferface with the
probabilistic programming language Stan (Carpenter et al. 2017, RStan
version 2.18.2 ), we also ran a Bayesian analysis
with a main effect of baseline interval and main effects of and
interactions between experimental condition and scalp topography. For
the baseline interval, we used a Student's \(t\) prior with three
degrees of freedom, centered at +1 and variance equal to 0.001. This
leads to a very sharp spike centered at 1 with heavy tails -- in more
casual terms, this is a very strong belief in traditional baseline
correction with nonetheless a willingness to change given enough
evidence. For the condition and topographical factors, we used normal
priors centered at 0 and with standard deviation equal to 2. This is
equivalent to the assumption that 60\% of effects are smaller than ±2µV
of 0 and 95\% effects are smaller than ±4µV, which is a reasonable ``no
outrageous'' effects assumption for language-related ERPs.

\begin{figure}[!htb]
\centering
\includegraphics[width=0.5\textwidth]{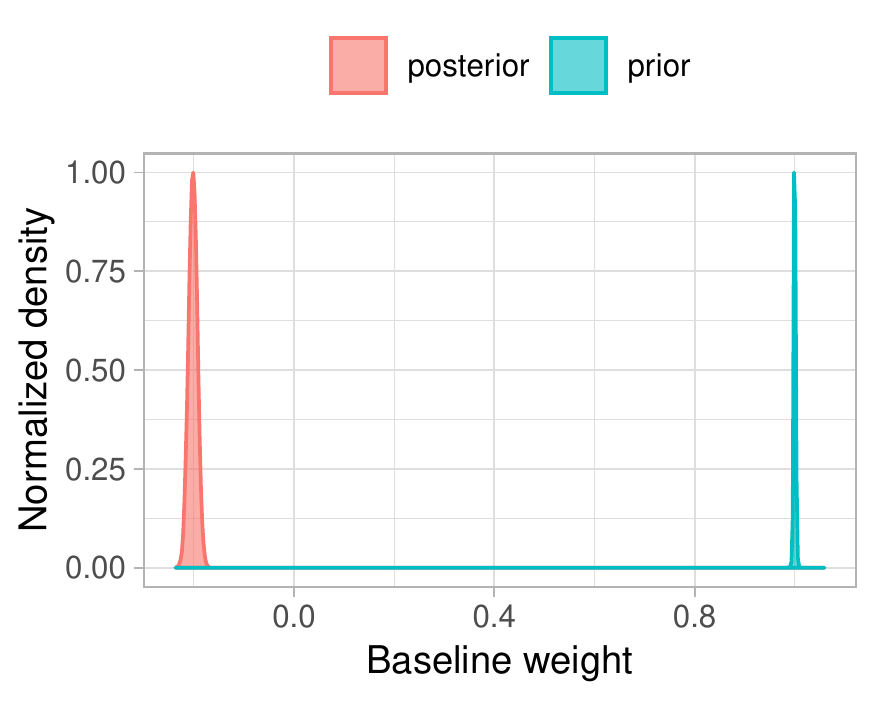}
\caption{Bayesian analysis of the appropriate weighting of the baseline interval.
         The prior (greenish blue color) places strong credibility in traditional baseline correction (narrow modal peak at +1), but has heavy tails that allow re-assigning credibility in light of sufficient evidence.
         The posterior (orange-ish red color) places most of its credibility at a small value in the direction opposite the traditional correction.
         Density normalized such that the maximum density within each distribution is 1.0.
         Model fit with 4 chains, each with 5\,000 post warm-up iterations; $\hat{R} = 1$ and $n_\text{eff} > 7\,000$ for all parameters.}
\label{fig:bayes}
\end{figure}

Figure \ref{fig:bayes} presents the resultant change in beliefs about
the correct weighting for the baseline interval. Even starting from such
a strong assumption, the posterior distribution still clearly places the
most credibility on a small, yet non-zero weighting for the baseline
interval in the direction \emph{opposite} the traditional direction.

\hypertarget{conclusion}{
\section{Conclusion}\label{conclusion}}

Baseline correction is in many ways the twin of filtering in EEG
preprocessing, serving both to replace stronger filtering and ultimately
functioning as a filter itself (see above discussion in
\emph{Psychophysiology} and \emph{Journal of Neuroscience Methods}).
However, traditional baseline correction is self defeating, increasing
noise and not affecting signal in exactly those situations fulfilling
its assumptions. Here, we have presented a straightforward extension of
the modern statistical analysis that supercedes the traditional baseline
correction, allowing the data to dynamically determine the strength of
the correction, while including both traditional baseline correction and
no baseline correction as limiting cases. Extending Tanner and
colleagues' comments a bit (2016), we can find out whether and how much
baseline correction is a good idea.

\hypertarget{acknowledgements}{
\section{Acknowledgements}\label{acknowledgements}}

Johanna Tromp and David Peeters kindly shared their data with me for the
practical example. I am grateful to several colleagues who have
commented on various drafts of this proposal in both casual conversation
and written form, including but not limited to Ingmar Brilmayer,
Franziska Kretzschmar, Benedikt Ehinger, Florian Hintz, Greta Kaufeld,
Suzanne Jongman, Darren Tanner and Andreas Widmann. In particular, I am
thankful to Andrea Martin for suggesting the current (sub)title. Two
anonymous reviewers provided invaluable feedback on theoretical and
practical aspects. All remaining mistakes are my own.

\clearpage

\hypertarget{references}{
\section*{References}\label{references}}
\addcontentsline{toc}{section}{References}

\hypertarget{refs}{}
\leavevmode\hypertarget{ref-akaike1974a}{}
Akaike, Hirotugu. 1974. ``A New Look at the Statistical Model
Identification.'' \emph{IEEE Transactions on Automatic Control} 19 (6):
716--23. \url{https://doi.org/10.1109/TAC.1974.1100705}.

\leavevmode\hypertarget{ref-baayen.vasishth.etal:2017jml}{}
Baayen, Harald, Shravan Vasishth, Reinhold Kliegl, and Douglas Bates.
2017. ``The Cave of Shadows: Addressing the Human Factor with
Generalized Additive Mixed Models.'' \emph{Journal of Memory and
Language} 94: 206--34. \url{https://doi.org/10.1016/j.jml.2016.11.006}.

\leavevmode\hypertarget{ref-baayendavidsonbates2008a}{}
Baayen, R. H., D. J. Davidson, and D. M. Bates. 2008. ``Mixed-Effects
Modeling with Crossed Random Effects for Subjects and Items.''
\emph{Journal of Memory and Language} 59: 390--412.

\leavevmode\hypertarget{ref-barrlevyscheepers2013a}{}
Barr, Dale J., Roger Levy, Christoph Scheepers, and Harry J. Tily. 2013.
``Random Effects Structure for Confirmatory Hypothesis Testing: Keep It
Maximal.'' \emph{Journal of Memory and Language} 68: 255--78.
\url{https://doi.org/10.1016/j.jml.2012.11.001}.

\leavevmode\hypertarget{ref-bates.maechler.etal:2015}{}
Bates, Douglas, Martin Maechler, Benjamin M. Bolker, and Steven Walker.
2015. ``Fitting Linear Mixed-Effects Models Using Lme4.'' \emph{Journal
of Statistical Software} 67 (1): 1--48.
\url{https://doi.org/10.18637/jss.v067.i01}.

\leavevmode\hypertarget{ref-buttonioannidismokrysz2013a}{}
Button, Katherine S, John P A Ioannidis, Claire Mokrysz, Brian A Nosek,
Jonathan Flint, Emma S J Robinson, and Marcus R Munafò. 2013. ``Power
Failure: Why Small Sample Size Undermines the Reliability of
Neuroscience.'' \emph{Nat Rev Neurosci}.
\url{https://doi.org/10.1038/nrn3475}.

\leavevmode\hypertarget{ref-buerkner2017joss}{}
Bürkner, Paul-Christian. 2017. ``Brms: An R Package for Bayesian
Multilevel Models Using Stan.'' \emph{Journal of Statistical Software}
80 (1). Foundation for Open Access Statistic.
\url{https://doi.org/10.18637/jss.v080.i01}.

\leavevmode\hypertarget{ref-carpenter.etal2017joss}{}
Carpenter, Bob, Andrew Gelman, Matthew D. Hoffman, Daniel Lee, Ben
Goodrich, Michael Betancourt, Marcus Brubaker, Jiqiang Guo, Peter Li,
and Allen Riddell. 2017. ``Stan: A Probabilistic Programming Language.''
\emph{Journal of Statistical Software} 76 (1). Foundation for Open
Access Statistic. \url{https://doi.org/10.18637/jss.v076.i01}.

\leavevmode\hypertarget{ref-clark1973a}{}
Clark, Herbert H. 1973. ``The Language-as-Fixed-Effect Fallacy: A
Critique of Language Statistics in Psychological Research.''
\emph{Journal of Verbal Learning and Verbal Behavior} 12: 335--59.
\url{https://doi.org/10.1016/S0022-5371(73)80014-3}.

\leavevmode\hypertarget{ref-cumming2014a}{}
Cumming, Geoff. 2014. ``The New Statistics: Why and How.''
\emph{Psychological Science} 20 (10): 1--23.
\url{https://doi.org/10.1177/0956797613504966}.

\leavevmode\hypertarget{ref-froemer.etal:2018fin}{}
Frömer, Romy, Martin Maier, and Rasha Abdel Rahman. 2018. ``Group-Level
Eeg-Processing Pipeline for Flexible Single Trial-Based Analyses
Including Linear Mixed Models.'' \emph{Frontiers in Neuroscience} 12.
Frontiers Media SA. \url{https://doi.org/10.3389/fnins.2018.00048}.

\leavevmode\hypertarget{ref-gaspar.etal2011N}{}
Gaspar, Carl M., Guillaume A. Rousselet, and Cyril R. Pernet. 2011.
``Reliability of ERP and Single-Trial Analyses.'' \emph{NeuroImage} 58
(2). Elsevier BV: 620--29.
\url{https://doi.org/10.1016/j.neuroimage.2011.06.052}.

\leavevmode\hypertarget{ref-gelmancarlin2014a}{}
Gelman, Andrew, and John Carlin. 2014. ``Beyond Power Calculations:
Assessing Type S (Sign) and Type M (Magnitude) Errors.''
\emph{Perspectives on Psychological Science}.
\url{https://doi.org/10.1177/1745691614551642}.

\leavevmode\hypertarget{ref-gramfort.luessi.etal:2013fn}{}
Gramfort, Alexandre, Martin Luessi, Eric Larson, Denis A Engemann,
Daniel Strohmeier, Christian Brodbeck, Roman Goj, et al. 2013. ``MEG and
Eeg Data Analysis with Mne-Python.'' \emph{Front Neurosci} 7: 267.
\url{https://doi.org/10.3389/fnins.2013.00267}.

\leavevmode\hypertarget{ref-green.macleod:2016mee}{}
Green, Peter, and Catriona J. MacLeod. 2016. ``SIMR: An R Package for
Power Analysis of Generalized Linear Mixed Models by Simulation.''
\emph{Methods in Ecology and Evolution} 7 (4): 493--98.
\url{https://doi.org/10.1111/2041-210X.12504}.

\leavevmode\hypertarget{ref-haukdavisford2006a}{}
Hauk, O., M. H. Davis, M. Ford, F. Pulvermüller, and W. D.
Marslen-Wilson. 2006. ``The Time Course of Visual Word Recognition as
Revealed by Linear Regression Analysis of Erp Data.'' \emph{NeuroImage}
30 (4): 1383--1400.
\url{https://doi.org/10.1016/j.neuroimage.2005.11.048}.

\leavevmode\hypertarget{ref-hoenig.heisey:2001as}{}
Hoenig, John M., and Dennis M. Heisey. 2001. ``The Abuse of Power: The
Pervasive Fallacy of Power Calculations for Data Analysis.'' \emph{The
American Statistician} 55 (1): 19--24.

\leavevmode\hypertarget{ref-judd.westfall.etal:2012pp}{}
Judd, Charles M., Jacob Westfall, and David A. Kenny. 2012. ``Treating
Stimuli as a Random Factor in Social Psychology: A New and Comprehensive
Solution to a Pervasive but Largely Ignored Problem.'' \emph{J Pers Soc
Psychol} 103 (1): 54--69. \url{https://doi.org/10.1037/a0028347}.

\leavevmode\hypertarget{ref-kruschke.liddell:2017pbr}{}
Kruschke, John K., and Torrin M. Liddell. 2017. ``The Bayesian New
Statistics: Hypothesis Testing, Estimation, Meta-Analysis, and Power
Analysis from a Bayesian Perspective.'' \emph{Psychonomic Bulletin \&
Review}, 1--29. \url{https://doi.org/10.3758/s13423-016-1221-4}.

\leavevmode\hypertarget{ref-lau.stroud.etal:2006bl}{}
Lau, Ellen, Clare Stroud, Silke Plesch, and Colin Phillips. 2006. ``The
Role of Structural Prediction in Rapid Syntactic Analysis.'' \emph{Brain
Lang} 98 (1): 74--88. \url{https://doi.org/10.1016/j.bandl.2006.02.003}.

\leavevmode\hypertarget{ref-luck2005a}{}
Luck, Steven J. 2005. \emph{An Introduction to the Event-Related
Potential Technique}. Cambridge, MA: MIT Press.

\leavevmode\hypertarget{ref-maess.schroger.etal:2016jnma}{}
Maess, B., E. Schröger, and A. Widmann. 2016. ``High-Pass Filters and
Baseline Correction in M/Eeg Analysis--Continued Discussion.''
\emph{Journal of Neuroscience Methods}, --.
\url{https://doi.org/10.1016/j.jneumeth.2016.01.016}.

\leavevmode\hypertarget{ref-maess.schroger.etal:2016jnm}{}
Maess, Burkhard, Erich Schröger, and Andreas Widmann. 2016. ``High-Pass
Filters and Baseline Correction in M/Eeg Analysis. Commentary on: `How
Inappropriate High-Pass Filters Can Produce Artefacts and Incorrect
Conclusions in Erp Studies of Language and Cognition'.'' \emph{Journal
of Neuroscience Methods} 266 (June). Elsevier BV: 164--65.
\url{https://doi.org/10.1016/j.jneumeth.2015.12.003}.

\leavevmode\hypertarget{ref-matuschek.kliegl.etal:2017jml}{}
Matuschek, Hannes, Reinhold Kliegl, Shravan Vasishth, Harald Baayen, and
Douglas Bates. 2017. ``Balancing Type I Error and Power in Linear Mixed
Models.'' \emph{Journal of Memory and Language} 94: 305--15.
\url{https://doi.org/10.1016/j.jml.2017.01.001}.

\leavevmode\hypertarget{ref-pernetsajdarousselet2011a}{}
Pernet, Cyril R, Paul Sajda, and Guillaume A Rousselet. 2011.
``Single-Trial Analyses: Why Bother?'' \emph{Frontiers in Psychology} 2
(322). \url{https://doi.org/10.3389/fpsyg.2011.00322}.

\leavevmode\hypertarget{ref-sassenhagen.alday:2016bl}{}
Sassenhagen, Jona, and Phillip M. Alday. 2016. ``A Common Misapplication
of Statistical Inference: Nuisance Control with Null-Hypothesis
Significance Tests.'' \emph{Brain and Language} 162: 42--45.
\url{https://doi.org/10.1016/j.bandl.2016.08.001}.

\leavevmode\hypertarget{ref-smithkutas2014b}{}
Smith, Nathaniel J, and Marta Kutas. 2014a. ``Regression-Based
Estimation of ERP Waveforms: II. Nonlinear Effects, Overlap Correction,
and Practical Considerations.'' \emph{Psychophysiology}.
\url{https://doi.org/10.1111/psyp.12320}.

\leavevmode\hypertarget{ref-smithkutas2014a}{}
---------. 2014b. ``Regression-Based Estimation of ERP Waveforms: I. The
rERP Framework.'' \emph{Psychophysiology}.
\url{https://doi.org/10.1111/psyp.12317}.

\leavevmode\hypertarget{ref-szucs.ioannidis2017Pb}{}
Szucs, Denes, and John P. A. Ioannidis. 2017. ``Empirical Assessment of
Published Effect Sizes and Power in the Recent Cognitive Neuroscience
and Psychology Literature.'' Edited by Eric-Jan Wagenmakers. \emph{PLOS
Biology} 15 (3). Public Library of Science (PLoS): e2000797.
\url{https://doi.org/10.1371/journal.pbio.2000797}.

\leavevmode\hypertarget{ref-tanner.morgan-short.etal:2015p}{}
Tanner, Darren, Kara Morgan-Short, and Steven J Luck. 2015. ``How
Inappropriate High-Pass Filters Can Produce Artifactual Effects and
Incorrect Conclusions in ERP Studies of Language and Cognition.''
\emph{Psychophysiology} 52 (8): 997--1009.
\url{https://doi.org/10.1111/psyp.12437}.

\leavevmode\hypertarget{ref-tanner.norton.etal:jnm}{}
Tanner, Darren, James J. S. Norton, Kara Morgan-Short, and Steven J.
Luck. 2016. ``On High-Pass Filter Artifacts (They're Real) and Baseline
Correction (It's a Good Idea) in Erp/Ermf Analysis.'' \emph{Journal of
Neuroscience Methods}, --.
\url{https://doi.org/10.1016/j.jneumeth.2016.01.002}.

\leavevmode\hypertarget{ref-tremblay.newman:2015p}{}
Tremblay, Antoine, and Aaron J. Newman. 2015. ``Modeling Nonlinear
Relationships in Erp Data Using Mixed-Effects Regression with R
Examples.'' \emph{Psychophysiology} 52: 124--39.
\url{https://doi.org/10.1111/psyp.12299}.

\leavevmode\hypertarget{ref-tromp.etal:2017brm}{}
Tromp, Johanne, David Peeters, Antje S. Meyer, and Peter Hagoort. 2017.
``The Combined Use of Virtual Reality and Eeg to Study Language
Processing in Naturalistic Environments.'' \emph{Behavior Research
Methods}. \url{https://doi.org/10.3758/s13428-017-0911-9}.

\leavevmode\hypertarget{ref-urbach.kutas:2006bp}{}
Urbach, Thomas P., and Marta Kutas. 2006. ``Interpreting Event-Related
Brain Potential (Erp) Distributions: Implications of Baseline Potentials
and Variability with Application to Amplitude Normalization by Vector
Scaling.'' \emph{Biological Psychology} 72 (3): 333--43.
\url{https://doi.org/10.1016/j.biopsycho.2005.11.012}.

\leavevmode\hypertarget{ref-widmann.schroger.etal:2015jnm}{}
Widmann, Andreas, Erich Schröger, and Burkhard Maess. 2015. ``Digital
Filter Design for Electrophysiological Data -- a Practical Approach.''
\emph{Journal of Neuroscience Methods} 250: 34--46.
\url{https://doi.org/10.1016/j.jneumeth.2014.08.002}.

\end{document}